%% file: main.tex
\documentclass[manuscript]{acmart}
\AtBeginDocument{%
  }

\setcopyright{acmlicensed}
\copyrightyear{2018}
\acmYear{2018}
\acmDOI{XXXXXXX.XXXXXXX}

\acmJournal{JACM}
\acmVolume{37}
\acmNumber{4}
\acmArticle{111}
\acmMonth{8}

\input{macros}

\begin{document}

\title{\ours{}: Enhancing LLM-based Recommendations with Attention-guided RAG from Web}

\author{Zihuai Zhao}
\affiliation{%
  \institution{The Hong Kong Polytechnic University}
  \country{Hong Kong SAR}
}
\email{zihuai.zhao@connect.polyu.hk}

\author{Yujuan Ding}
\affiliation{%
  \institution{The Hong Kong Polytechnic University}
  \country{Hong Kong SAR}
}
\email{dingyujuan385@gmail.com}

\author{Wenqi Fan}
\authornote{Corresponding author.}
\email{wenqifan03@gmail.com}
\author{Qing Li}
\email{qing-prof.li@polyu.edu.hk}
\affiliation{%
  \institution{The Hong Kong Polytechnic University}
  \country{Hong Kong SAR}
}

\renewcommand{\shortauthors}{Zhao et al.}

\begin{abstract}
Recommender systems play a vital role in alleviating information overload and enriching users' online experience.
In the era of large language models (LLMs), LLM-based recommender systems have emerged as a prevalent paradigm for advancing personalized recommendations.
Recently, retrieval-augmented generation (RAG) has drawn growing interest to facilitate the recommendation capability of LLMs, incorporating useful information retrieved from external knowledge bases.
However, as a rich source of up-to-date information, the web remains under-explored by existing RAG-based recommendations.
In particular, unique challenges are posed from two perspectives: one is to generate effective queries for web retrieval, considering the inherent knowledge gap between web search and recommendations; another challenge lies in harnessing online websites that contain substantial noisy content.
To tackle these limitations, we propose \textbf{\ours{}}, a novel web-based RAG framework, which takes advantage of the reasoning capability of LLMs to interpret recommendation tasks into queries of user preferences that cater to web retrieval.
Moreover, given noisy web-retrieved information, where relevant pieces of evidence are scattered far apart, an insightful \adapter{} is designed to enhance LLM attentions between distant tokens of relevant information via message passing.
Extensive experiments have been conducted to demonstrate the effectiveness of our proposed web-based RAG methods in recommendation scenarios.
\end{abstract}




\maketitle

\input{01introcduction}

\input{02relatedwork}

\input{03method}

\input{04experiment}

\input{05conclusion}


\bibliographystyle{ACM-Reference-Format}
\bibliography{reference}

%








\end{document}

%% file: macros.tex
\usepackage[dvipsnames]{xcolor}
\usepackage{subfigure}
\usepackage{tabularx}
\usepackage{threeparttable}
\usepackage{makecell}
\usepackage{bm}
\usepackage{colortbl}
\usepackage{bbold}
\usepackage{natbib}
\usepackage{enumitem}
\usepackage[most]{tcolorbox}
\usepackage{multirow}
\usepackage{CJKutf8}

\definecolor{atomictangerine}{rgb}{1.0, 0.6, 0.4}
\definecolor{ballblue}{rgb}{0.13, 0.67, 0.8}

\definecolor{color1}{HTML}{f0e8d0}
\definecolor{color2}{HTML}{e6e7f6}
\definecolor{color3}{HTML}{e6f6f5}
\newcommand{\highlighta}[1]{{\colorbox{color1}{#1}}}
\newcommand{\highlightb}[1]{{\colorbox{color2}{#1}}}
\newcommand{\highlightc}[1]{{\colorbox{color3}{#1}}}

\usepackage{pifont}
\newcommand{\yessymbol}{\color{green!60!black}\ding{51}}
\newcommand{\nosymbol}{\color{red!60!black}\ding{55}}

\usepackage[utf8]{inputenc}
\usepackage[T1]{fontenc} 
\usepackage{listings}
\newcommand{\fon}[1]{\fontfamily{#1}\selectfont}

\newcommand{\llmrec}{{LLM-RS}}
\newcommand{\ours}{{WebRec}}
\newcommand{\adapter}{{MP-Head}}

\definecolor{slateblue}{RGB}{106, 90, 205}

%% file: 01introcduction.tex
\section{INTRODUCTION}
Large language models (LLMs) have achieved remarkable breakthroughs in advancing next-generation recommender systems, namely LLM-based Recommender Systems (\llmrec{})~\cite{zhao2024recommender}.
In particular, LLMs equipped with billion-scale parameters have exhibited remarkable language understanding and reasoning abilities~\cite{zhao2023survey}. 
These capabilities enable LLMs to effectively capture diverse user preferences by leveraging rich textual side information in recommender systems, such as user profiles and item descriptions.
However, LLMs often struggle when faced with knowledge-intensive queries, as their pre-trained knowledge may be incomplete or outdated. This leads to LLMs providing inaccurate recommendations, such as hallucinated content (e.g., non-existent products like ``iPhone 9'') and factual inaccuracies in real-world scenarios~\cite{wang2024speculative}.
To address these limitations, retrieval-augmented generation (RAG) has emerged as a promising technique that bridges retrieval systems and LLMs, incorporating coherent and informative knowledge before generating responses.
In other words, LLMs equipped with RAG modules can leverage external knowledge bases, retrieving reliable and relevant information as augmented inputs of LLMs~\cite{wang2025knowledge, fan2024survey, ning2025retrieval}.

\begin{figure*}[t]
    \centering
    \includegraphics[width=\textwidth]{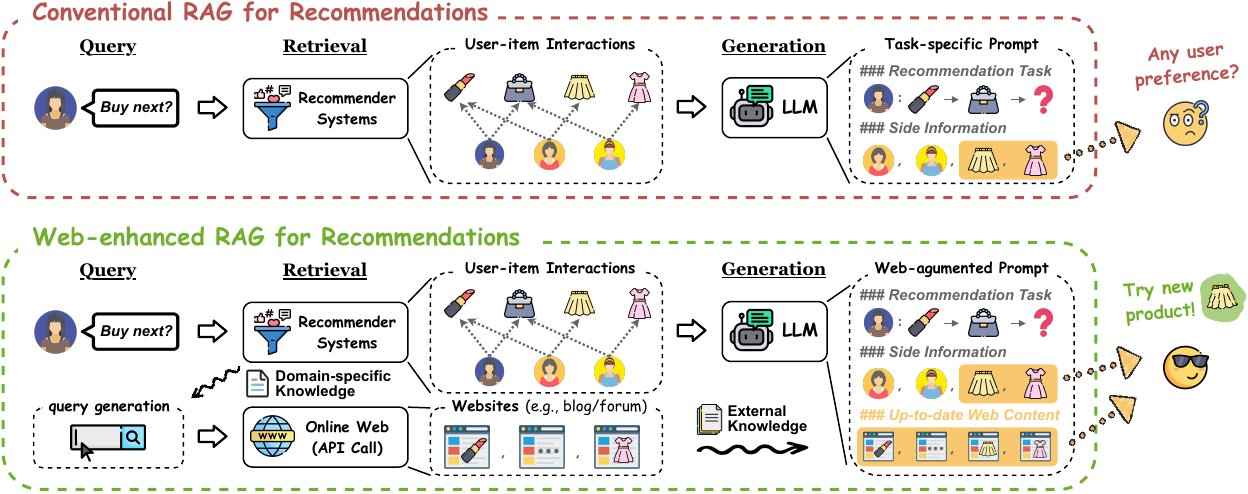}
    \caption{
    Illustration of web-enhanced RAG for recommendations.
    In addition to domain-specific knowledge from recommender systems, online websites offer a distinct advantage by providing access to up-to-date data as external knowledge.
    This contributes to fulfilling the timely information needs in recommendations, such as the latest customer feedback online, to facilitate the understanding of user preferences.
    }
\vskip -0.15in
\label{fig:motivation}  
\end{figure*}

Recently, the integration of web search and LLMs via RAG has drawn growing interest from existing studies.
Compared to static knowledge databases that require periodic updates, the web provides access to up-to-date data to fulfill timely information needs.
Notably, the web harnesses search engines that are facilitated with advanced information retrieval capabilities after decades of development, enabling the web to serve as a rich source of external knowledge in RAG systems to retrieve online websites based on queries.
For example, HtmlRAG~\cite{tan2025htmlrag} and WebWalker~\cite{wu2025webwalker} have proposed web search and result pruning pipelines, enhancing the generation quality of LLMs in QA tasks.
Building upon the recent advances in web-based RAG, new opportunities have been uncovered to facilitate the performance of LLM-based recommendations.
As illustrated in Fig.~\ref{fig:motivation}, the web offers a distinct advantage by augmenting the domain-specific knowledge with large-scale and up-to-date websites, where relevant external knowledge can be retrieved to facilitate the understanding of user preferences. 
For example, when \llmrec{} lacks pre-trained knowledge or side information to fulfill accurate recommendations, online websites that aggregate public reviews (e.g., personal blogs and discussion forums) can provide valuable information to encode item features or understand diverse user preferences for personalized recommendations.

Despite the great potential of web-based RAG in facilitating \llmrec{}, online websites often contain substantial noisy content~\cite{deng2023mind2web, xie2024weknow}, such as long-context conversations and misleading content, making it challenging to extract high-quality web data that meets the needs of recommendation tasks.
However, most existing works on RAG-enhanced \llmrec{} are limited to matching-based methods, where user/item IDs are available for accurate retrieval, such as similar user profiles~\cite{wu2024coral, kemper2024retrieval} and interacted item reviews~\cite{salemi2024optimization} in recommendation metadata.
Due to the significant gap between web content and recommendations in terms of the data structure and scope, these matching-based RAG potentially fall short in identifying useful relevant information over noisy web data.
For example, when searching for movie recommendations, the retrieved websites may contain lengthy forum discussions, advertisements, or vague personal blogs that implicitly reflect user preferences. 
This leads to existing RAG paradigms hardly capturing the information needs of \llmrec{}, such as understanding nuanced user preferences and distinguishing between conflicting opinions for accurate recommendations.
Therefore, it is imperative to delve into the design of task-specific RAG to facilitate recommendation performance over noisy web information.

Typically, the RAG pipeline involves two stages: \textit{retrieval} (e.g., external knowledge bases) and \textit{generation} (e.g., LLM responses). To facilitate the recommendation capability of LLMs with web-based RAG, unique challenges are posed from two perspectives as follows:
\begin{itemize}[leftmargin=*, noitemsep, topsep=0pt]
    \item \textbf{Retrieval stage:}
    Effective web retrieval in RAG requires specific and concise queries to retrieve information that contributes to enhancing the generation process.
    Such requirements can be easily fulfilled in general QA tasks by directly using questions as queries, such as \textit{``Who is the president of America?''}.
    However, given a recommendation query of user-item interactions and asking \textit{``What will the user buy next?''}, less useful results are likely to be retrieved since the web inherently lacks the capability to address recommendation problems.
    This leads to query generation in most existing web-based RAG methods that directly apply task keywords as the retrieval query~\cite{yan2024corrective, tan2025htmlrag}, becoming impractical.
    Therefore, the design of adaptive recommendation queries that cater to web retrieval is required.
    
    \item \textbf{Generation stage:} 
    \llmrec{} typically relies on textual prompts, such as the description of recommendation tasks, to generate accurate recommendations by capturing the semantic correlation between prompt tokens in recommendation tasks.
    However, online websites often contain substantial noisy content, leading to a large volume of irrelevant information (e.g., long-context conversations) for LLMs to handle.
    Such noisy information manifests as long-distance semantic dependencies, where relevant pieces of evidence are scattered far apart in the textual prompt.
    This leads \llmrec{} struggling to capture distant correlations between web content and recommendation tasks, which weakens its ability to reason about user preferences.
    Consequently, directly training \llmrec{} on web-augmented data, which contains substantial noisy content, can fail to address accurate recommendations.
\end{itemize}

\noindent
Related to our web-based RAG task for recommendations, recent studies on LLM-based retrieval have discovered \textit{retrieval heads} in LLM attention mechanisms, which are chiefly responsible for retrieving relevant information over long-context inputs~\cite{wu2024retrieval, zhou2024role, su2024dragin}.
In particular, \citet{wu2024retrieval} have pointed out that attention heads are capable of redirecting relevant information tokens from input to output in the Needle-in-a-Haystack test~\cite{kamradt2023needle} (i.e., copy-paste tasks).
Another study SnapKV~\cite{li2024snapkv} has also demonstrated the functionality of attention heads to identify the essential tokens that carry task-relevant knowledge over long-context information.
These advanced findings shed light on an effective approach to capture long-distance dependencies between noisy web content and recommendation tasks by utilizing the attention feature.
Intuitively, the attention feature of relevant tokens that contain critical information in recommendation tasks, such as recommendation prompt tokens, can serve as a latent similarity measurement to guide the retrieval of task-relevant information over noisy web content.

In light of our inspiration for facilitating \llmrec{} via attention-guided web retrieval, we proposed a novel framework, namely \ours{}, to address the unique challenges of bridging web retrieval and LLM-based recommendations.
In the retrieval stage, we take advantage of the reasoning capability of LLMs to interpret recommendation tasks (e.g., \textit{``Recommend the next item''}) into specific and informative queries, such as the detailed description of user preferences, that cater to web retrieval.
Specifically, we first apply LLMs to generate token sequences, and then sample critical tokens as high-quality retrieval queries based on carefully designed scoring of the LLM information needs in recommendation tasks.
Notably, our retrieval method can directly work on recommendation prompts, alleviating the necessity of fine-tuning LLMs or massive prompt engineering for retrieval tasks.
In the generation stage, we draw inspiration from the unique retrieval capability of attention mechanisms and develop a novel MP-Head to enhance LLM attentions for modeling long-distance dependencies via message passing. 
In particular, the key insight behind message passing lies in regarding textual prompts as a token-level graph~\cite{huang2020aligned}, where entities represent the attention features of tokens and relations are modeled by learnable features.
During message passing, the distant correlations between noisy web content and recommendation tasks can be effectively modeled by the \textit{relation} between \textit{entities}, enhancing the attention between distant tokens as one-hop connectivity. 
Notably, the MP-Head serves as an additional attention head, which can be seamlessly integrated into LLM layers in an adapter manner.
Therefore, in addition to the semantic correlation captured by vanilla attention mechanisms, the learnable correlation via message passing helps capture relevant information over noisy web content to facilitate recommendation performance.
In summary, the main contributions of this paper are organized as follows:
\begin{itemize}[leftmargin=*, noitemsep, topsep=0pt]
    \item We propose a novel framework named \ours{} that retrieves up-to-date information from online websites to facilitate LLM-based recommendations via web-based RAG.
    Notably, our method can directly work on recommendation prompts by capturing the information needs of LLMs as retrieval queries, without fine-tuning or massive prompt engineering for retrieval tasks.

    \item We investigate the significant knowledge gap between web search and recommender systems, where noisy web data hardly contributes to the recommendation capability of LLMs.
    Drawing inspiration from the retrieval capability of the attention mechanism in LLMs, we introduce an attention-guided RAG approach, instead of existing matching-based methods.
    
    \item To address the long-distance dependencies between noisy web information and recommendation tasks, we design a novel \adapter{} that captures learnable correlations via message passing, enhancing LLM attentions between distant tokens as one-hop connectivity.
    In particular, the proposed \adapter{} can be seamlessly integrated into LLM layers in an adapter manner.

    \item Extensive experiments on different real-world recommendation datasets are conducted, demonstrating the effectiveness of our proposed methods under diverse sources of web information.
\end{itemize}

%% file: 02relatedwork.tex
\section{RELATED WORK}
In this section, we elaborate on the paradigm of LLM-based recommender systems and the emergence of retrieval-augmented generation in the domain of generative models.

\subsection{LLM-based Recommender Systems (\llmrec{})}
Recommender systems provide personalized suggestions tailored to user preferences, facilitating user experience across diverse applications, such as e-commerce, job matching, and social media platforms~\cite{fan2019graph, fan2020graph, liu2025score}.
Recently, Large Language Models (LLMs) have emerged as a prevalent paradigm for advancing personalized recommendations.
To seamlessly adapt LLMs into task-specific recommendations, many existing works utilize textual prompts to guide the language understanding of recommendation information, such as user-item interactions, harnessing the rich semantic knowledge of LLMs~\cite{zhao2024recommender}.
Notably, LLMs equipped with billion-scale parameters have demonstrated unprecedented language understanding and reasoning capabilities~\cite{lyu2025pla, zhao2025investigating}, which enables capturing diverse user preferences based on rich textual side information in recommender systems, such as user profiles and item descriptions.
For instance, ~\citet{qu2025generative} take advantage of the reasoning from LLMs to condition on positive and negative interactions, which contributes to learning a generative representation of future items, facilitating the understanding of user preferences.
As a result, the remarkable breakthroughs of \llmrec{} have drawn growing interest in recent studies.

\subsection{Retrieval-augmented Generation (RAG)}
Despite the unprecedented capabilities of \llmrec{}, LLMs often struggle when faced with knowledge-intensive queries, as their pre-trained knowledge may be incomplete or outdated.
This leads to LLMs providing inaccurate recommendations, such as hallucinated content (e.g., non-existent products like ``iPhone 9'') and factual inaccuracies in real-world scenarios~\cite{wang2025knowledge, wang2024speculative}.
To address such issues, RAG has emerged as a promising technique that bridges retrieval systems and LLMs, incorporating coherent and informative knowledge before generating responses.
For instance, LLMs equipped with RAG modules can leverage external knowledge bases, retrieving reliable and relevant information as augmented inputs of LLMs~\cite{fan2024survey, ning2025retrieval}.
To align retrievers with the varying knowledge preferences of LLMs, ~\citet{dong2025understand} introduce a preference knowledge construction pipeline, leveraging multiple query augmentation strategies that allow LLMs to capture knowledge tailored to their reasoning preferences.
Similarly, distinctive reranking strategies are proposed to rank the set of retrieved documents that accurately meet the demands of LLMs, such as RankCoT~\cite{wu2025rankcot}, CFRAG~\cite{shi2025retrieval}, and R2LLMs~\cite{zuccon2025r2llms}, highlighting the effective design of retrieval strategies in RAG systems, which contributes to accurately capturing the information needs of LLMs.

In the field of recommender systems, initial works have been conducted to facilitate the recommendation capability of LLMs with RAG techniques, harnessing the rich side information retrieved from recommender systems.
For instance, CoRAL~\cite{wu2024coral} aligns LLM reasoning with task-specific user–item interaction knowledge by retrieving collaborative evidence directly into prompts.
RA-Rec~\cite{kemper2024retrieval} utilizes comprehensive item reviews retrieved from known metadata to facilitate closer alignment with users’ potential interests.
ROPG~\cite{salemi2024optimization} attempts to optimize retrieval models that supply a constrained set of personal documents to LLMs, and enhance personalized recommendations.
However, retrieving up-to-date information from the web, which serves as a rich source of open-domain knowledge, remains under-explored and shows great potential to facilitate \llmrec{}.

%% file: 03method.tex
\section{METHODOLOGY}
In this section, we propose \ours{}, a web-based RAG framework to facilitate LLM-RS, in addressing the distinctive challenges in the retrieval and generation stages of the RAG pipeline.
\begin{itemize}[leftmargin=*, noitemsep, topsep=0pt]
    \item In the \textbf{retrieval stage}, an insightful training-free strategy is designed to bridge the knowledge gap between web search and recommendation tasks, as illustrated in Figure~\ref{fig:retrieval}.
    Notably, we take advantage of the reasoning capability of LLMs to interpret recommendation tasks into informative queries, such as the description of user preferences, that cater to web retrieval.
    In particular, we first apply LLMs to generate token sequences, and then sample critical tokens as high-quality retrieval queries based on carefully designed scoring of the LLM information needs in recommendation tasks.

    \item In the \textbf{generation stage}, web-augmented information often contains substantial noisy content, where relevant pieces of evidence are scattered far apart.
    This leads \llmrec{} struggling to capture distant correlations between web content and recommendation tasks, which weakens its ability to reason about user preferences.
    To tackle such limitations, we propose a novel \adapter{} to enhance LLM attentions between distant tokens via message passing.
    As illustrated in Figure~\ref{fig:method}, the long-distance dependencies learned by \adapter{}, which serves as an additional attention head in LLM layers, contribute to capturing relevant information over noisy web content to facilitate recommendation performance. 
\end{itemize}

\subsection{Retrieval Stage of \ours{}}
In the retrieval stage, we aim to search for relevant web information in recommendation tasks via API calls, such as Google and Bing APIs.
However, the web retrieval typically employs keyword matching between search queries and web data, which inherently lacks the capabilities of recommendation tasks, such as next-item predictions.
For example, the web could retrieve accurate information in QA tasks like \textit{``Who is the president of America?''}, while hardly capture the information need in recommendations like \textit{``What will the user buy next?''}.
Therefore, the design of adaptive recommendation queries that cater to web retrieval is required.
As a common solution, existing web-based RAG~\cite{wu2025webwalker, tan2025htmlrag} take advantage of the reasoning capability of LLMs to generate appropriate search queries in open-domain QA.
Given the significant knowledge gap between web search and recommendations, massive training data or dedicated prompt engineering can be required to benchmark the performance of LLM-generated queries.
To address these limitations, we propose to capture high-quality queries based on the information needs during the inference (i.e., training-free) of LLMs in recommendation tasks.

\begin{figure*}[t]
    \centering
    \includegraphics[width=\textwidth]{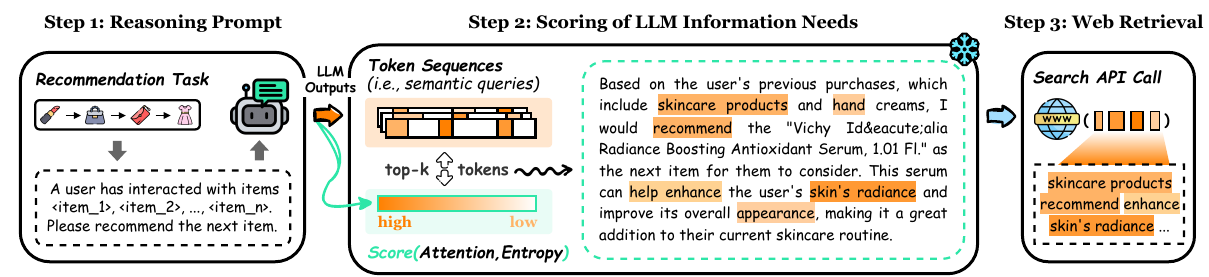}
    \caption{
    Overview of the \textbf{retrieval stage} of \ours{}.
    In step 1, we take advantage of the reasoning capability of LLMs to interpret recommendation tasks as semantic queries.
    In step 2, retrieval queries can be generated to address the information needs of LLMs by carefully scoring generated tokens, without fine-tuning or massive prompt engineering for retrieval tasks.
    In step 3, the web retrieval can be conducted via search APIs.
    }
\vskip -0.15in
\label{fig:retrieval}  
\end{figure*}

As illustrated in Figure~\ref{fig:retrieval}, we take advantage of the reasoning capability of LLMs to interpret recommendation tasks as semantic queries that cater to web retrieval.
Compared to document retrieval from conventional knowledge bases, the web equipped with search engines alleviates the design of retrieval tools, where relevant web content can be retrieved based on LLM-generated retrieval queries.
Specifically, we quantify the semantic importance of LLM-generated tokens to sample keywords as high-quality retrieval queries, since they represent the information need of LLMs in recommendation tasks.
Notably, our retrieval method can directly work on recommendation prompts, alleviating the necessity of fine-tuning LLMs or massive prompt engineering for retrieval tasks.
In the following sections, we elaborate on the implementation of each step in Figure~\ref{fig:retrieval}.

\subsubsection{\textbf{Step 1:} Reasoning Prompt}
Instead of requiring dedicated prompt engineering or task-specific fine-tuning (i.e., web retrieval tasks) for LLMs to directly generate search queries, we aim to capture query keywords from LLM outputs generated by recommendation prompts.
In other words, the reasoning capability of LLMs is leveraged to perform recommendation tasks, where available prompts can be obtained based on recommendation datasets as follows:

\begin{tcolorbox}[colback=white]
    Below is an instruction that describes a task. Please write a response that appropriately completes the request.
    
    \tcblower
    
    \textbf{[Instruction]}
    
    A user has bought $\langle \text{item}\_1 \rangle$, $\langle \text{item}\_2 \rangle$, ..., and $\langle \text{item}\_10 \rangle$ in the previous. Please recommend the next item for this user to buy from the following item title set: $\langle \text{candidate}\_1 \rangle$, $\langle \text{candidate}\_2 \rangle$, ..., and $\langle \text{candidate}\_20 \rangle$. The recommendation is
\end{tcolorbox}

\noindent It is worth noting that, compared to prompting LLMs for accurate query generation, the recommendation prompt alleviates task-specific requirements on the pre-trained knowledge of LLMs, such as output format and recommendation accuracy.
In particular, LLMs are asked to generate open-domain recommendations based on their reasoning capability, which usually includes the analysis of user preferences, examples of recommended items, and corresponding reasons.

\subsubsection{\textbf{Step 2:} Scoring of LLM Information Needs}
Following the recommendation prompt, we take advantage of the remarkable reasoning capability of LLMs to interpret recommendation knowledge as semantic queries that cater to web retrieval.
In particular, the essential information required in recommendations can be captured by the LLM's information needs during generation (i.e., textual outputs), which serves as high-quality queries for web retrieval.
To quantify the information needs of LLMs in recommendations, we integrate the semantic importance with the generation confidence of LLM outputs.
These combinations contribute to ranking the keywords of LLM-generated recommendations while ignoring tokens with high generation confidence, which indicate minimal necessity for retrieving external knowledge.

Building upon the above insights, we first utilize the attention score of generated tokens to calculate their semantic importance, which measures the influence of each token on subsequent tokens.
Consider a sequence of recommendation outputs generated by LLMs, denoted as $\mathcal{T} = \{t_1, t_2, ..., t_n \}$, the attention scores are calculated by the dot product between the query vectors Q of each token $t_i$ and the key vectors $K$ of all other tokens $t_j$ as follows:
\begin{equation}
    A_{i, j} = \bm{\mathrm{softmax}}(\frac{Q_i K_j^T}{\sqrt{d_k}}) \quad \forall 1 \leq i, j \leq n,
\end{equation}
where the matrix of attention scores $A \in \mathbb{R}^{n*n}$ and $d_k$ is the dimension of the key and query vectors.
To address the semantic importance of each generated token, we measure the attention scores over succeeding tokens that serve as its influence on recommendation outputs, since LLM generation is determined by the interpretation of preceding context.
Formally, given each token $t_i \in \mathcal{T}$, we quantify its semantic importance by calculating the maximum attention during the generation of all succeeding tokens $t_{j>i}$, thereby the score of semantic importance can be given by:
\begin{equation}
    s_i^{\mathrm{attention}} = \mathop{\max}_{j>i}A_{j, i}.
\end{equation}
Subsequently, we aim to combine the semantic importance of tokens with their generation confidence to score the information need of LLMs in recommendations, such as the keywords of generated tokens.
To score generation confidence at the token level, \ours{} quantifies the uncertainty of LLM about its own generated recommendation outputs.
In particular, LLM's uncertainty can be measured by the entropy of the probability distribution assigned to each generated token among the full vocabulary, thereby the score of generation confidence is formulated as:
\begin{equation}
    s_i^{\mathrm{entropy}} = -\mathop{\sum}_{v \in \mathcal{V}}P_i(v) \mathrm{log}P_i(v).
\end{equation}
where $P_i$ is the probability distribution of generating token $t_i$ that $t_i = \mathop{\arg\max}_v P_i(v)$, and $v \in \mathcal{V}$ denotes any token in the vocabulary of LLMs.
Finally, a comprehensive score in addressing the information needs of LLMs in recommendations can be computed by the combination of semantic importance and generation confidence as follows:
\begin{equation}
    s_i =  s_i^{\mathrm{attention}} \cdot s_i^{\mathrm{entropy}}.
\end{equation}
In other words, generated tokens with high scores indicate a large influence on generating recommendation outputs, while LLMs are relatively uncertain about their (i.e., tokens) generations. 
Therefore, we regard these tokens as high-quality keywords for retrieving external knowledge, since they represent the information needs of LLMs in recommendations.

\subsubsection{\textbf{Step 3:} Web Retrieval}
Following the scoring of semantic importance and generation confidence, retrieval keywords can be ranked from the generated tokens of LLMs.
Notably, compared to prompting LLMs to directly generate retrieval queries, our keyword-based methods can directly work on recommendation prompts.
This contributes to alleviating the necessity of fine-tuning LLMs or massive prompt engineering for retrieval tasks.
In particular, we aim to sample retrieval keywords from recommendation outputs (i.e., generated tokens) as high-quality retrieval queries.
However, the naive ranking of retrieval keywords may lead to diminishing the completeness of relevant information, such as the entire title of recommended items.
To address such limitations, \ours{} performs keyword clustering by pooling the scores of generated tokens, and then selects top-$k$ keywords as the query for web retrieval as follows:
\begin{equation}
    \{t_i\}_{i \in \mathcal{K}} \gets \mathrm{\bf topk}\{\mathrm{AvgPool}(s_1, s_2, ..., s_n)\},
\end{equation}
where $\mathcal{K}$ denotes the set of top-k indices based on the pooled scores.
Finally, the query for web retrieval can be constructed by concatenating the selected keywords, which indicate the information needs of LLMs in recommendations.
In other words, the web retrieval aims to provide relevant information from websites as additional prompts to address LLMs' information needs in recommendations, such as up-to-date public reviews.
For example, when LLMs lack pre-trained knowledge of newly released products, online websites that aggregate public reviews can provide valuable information on customer opinions and product performance.

\subsection{Generation Stage of \ours{}}
\begin{figure*}[t]
    \centering
    \includegraphics[width=\textwidth]{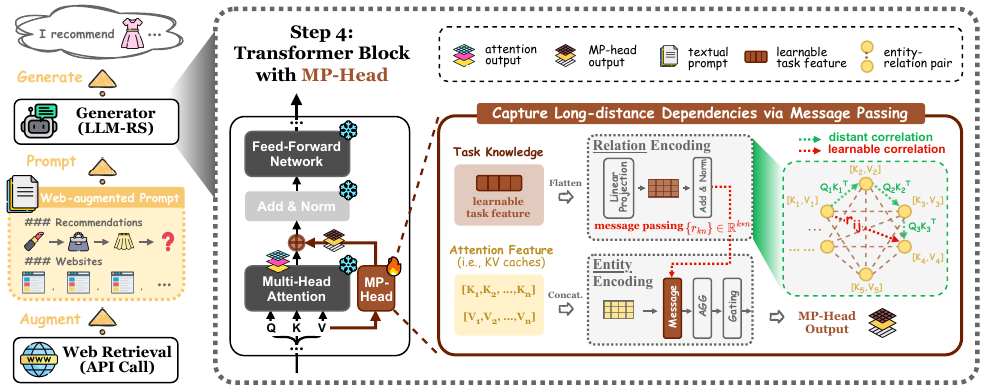}
    \caption{
    Overview of the \textbf{generation stage} of \ours{}.
    On the left, we illustrate the pipeline continued from web retrieval that \textit{\textbf{Step 1-3}} in Fig.~\ref{fig:retrieval}.
    In the right block, the proposed framework of LLM's transformer block with \adapter{} is presented, where \adapter{} takes the original attention features as inputs and models their long-distance dependencies via message passing.
    Notably, the learned \adapter{} output can be seamlessly integrated into the attention output to facilitate LLM-based recommendations over noisy web information.
    }
\vskip -0.15in
\label{fig:method}  
\end{figure*}

In the generation stage, \llmrec{} typically model the semantic correlation in textual prompts to generate task-specific recommendations.
For example, the probability of output tokens (e.g., item titles) can be measured by their semantic correlation to input tokens of recommendation data.
However, these semantic correlations might diminish significantly over noisy web information due to long-distance dependencies.
This leads to \llmrec{} hardly capture task-relevant correlations from the retrieved web information to facilitate recommendation performance.
In light of recent findings on attention heads that are responsible for retrieving information in LLMs~\cite{wu2024retrieval, zhang2024llama}, we aim to take advantage of attentions to model task-relevant correlations between web and recommendations.
As shown in Figure~\ref{fig:method}, we design a novel \textbf{\adapter{}} that captures long-distance correlations via message passing (MP), which can be seamlessly integrated into LLM attentions to effectively model task-relevant correlations over noisy web information.

\subsubsection{\textbf{Step 4:} Transformer Block with \adapter{}}
Given noisy web information, vanilla attention mechanisms capture the task-relevant information merely based on semantic correlations, which exhibit long-distance dependencies due to the large volume of irrelevant web content.
In particular, the attention models how much ``focus'' the information receives from \llmrec{} by calculating the dot product of textual token embeddings (i.e., semantic correlation).
Formally, each token embedding $x_i$ is first mapped into a vector pair $(Q_i, K_i, V_i)$ by LLMs, which is analogous to the concept of \textit{query}, \textit{key} and \textit{value} in retrieval systems~\cite{vaswani2017attention, guo2024attention}.
Subsequently, LLMs utilize a set of attention heads $\{\mathbf{head}_h\}_{h=1}^{N_{\mathrm{head}}}$ in transformer blocks to compute the attention at each token $i$, which can be largely affected by the relative distance between tokens as follows:
\begin{equation}\label{eq:attention}
    \mathbf{head}_h(x_i) \gets \mathrm{Attention}(Q_i, K_i, V_i) = \sum_{t \leq i}^i \bm{\mathrm{softmax}}(\frac{Q_i K_t^T}{\sqrt{d_k}})V_t,
\end{equation}
where $t$ denote any preceding token and $d_k$ is the dimensionality of \textit{query} and \textit{key} vectors.
Finally, all heads are concatenated to indicate the attention $\mathbf{a}_i$ from different representation subspaces, which can be formulated as:
\begin{equation}\label{eq:head}
    \mathbf{a}_i = \mathrm{CONCAT}(\mathbf{head}_1(x_i), \mathbf{head}_2(x_i), ..., \mathbf{head}_h(x_i)).
\end{equation}
To capture long-distance dependencies in recommendation tasks over web tokens, \adapter{} serves as an additional attention head that models task-relevant correlation between tokens with 1-hop distance via message passing.
In particular, these task correlations modeled by \adapter{} are seamlessly integrated with the semantic correlations captured by vanilla attention heads, which serve as attention features from an augmented representation subspace.
Notably, the proposed \adapter{} only introduces one dimension to the original LLM attention (i.e., $\mathbf{a}_i$) given by:
\begin{equation}\label{eq:mp-head}
     \mathbf{a}_i^{MP} = \mathrm{CONCAT}(\mathbf{head}_{MP}(x_i), \mathbf{a}_i),
\end{equation}
where $\mathbf{head}_{MP}$ (i.e., $\mathbf{a}_i^{MP}[0]$) denotes the output of \adapter{} that contributes to assigning higher attention scores to task-relevant tokens.
As shown in Figure~\ref{fig:method}, our proposed \adapter{} framework consists of three key components:
\begin{itemize}[leftmargin=*, noitemsep, topsep=0pt]
    \item \textbf{Entity Encoding}.
    In the multi-head attention of LLMs, the key-value (KV) embeddings of each token are cached by attention heads during generation.
    Therefore, to model the long-distance dependency between tokens, \adapter{} takes advantage of these KV features as \textit{entity} to represent each token.
    Compared to vanilla attention heads, which can be largely affected by token distances in Eq.~\eqref{eq:attention}, \adapter{} are capable of modeling long-distance dependencies by the connectivity between each token \textit{entity} with 1-hop distance, as shown in Fig.~\ref{fig:method}.

    \item \textbf{Relation Encoding}. By analogy with the semantic correlation between tokens captured by vanilla attention (i.e., dot product $QK^T$ of \textit{query}-\textit{key}), our \adapter{} adopts a learnable task feature as the \textit{query} to model task-relevant correlations between tokens. 
    Specifically, this task feature is obtained by the learnable parameters for fine-tuning \llmrec{}, such as prompt embeddings~\cite{liao2024llara,kim2024large, ji2024genrec} or trainable adapter parameters~\cite{ji2024genrec, geng2022recommendation, qu2024tokenrec} used in existing works.
    In other words, \adapter{} is not responsible for learning the task feature, but leveraging it to capture task-relevant correlations.
    Therefore, \adapter{} takes the task feature to encode the \textit{relation} between each token \textit{entity}, where the task-relevant correlation is modeled by their similarities.

    \item \textbf{Message Passing}. Given the representations of \textit{entity} and \textit{relation}, we can model the web-augmented prompt that includes both web and recommendation tokens as a prompt graph.
    Subsequently, message passing over the prompt graph is performed to update \textit{entity} representations, aggregating each token with other tokens (i.e., one-hop neighbors) based on their \textit{relation} representations.
    Thereby, task-relevant tokens over noisy web information can exhibit strong similarity (i.e., task-relevant correlation) with each other, which contributes to generating high attention scores between these tokens by \adapter{}. 
\end{itemize}
In the following sections, we elaborate on the implementation of these components to explain the generation of \adapter{} output $\mathbf{head}_{MP}$ defined in Eq.~\eqref{eq:mp-head}.

\subsubsection{Implementation of Entity Encoding}
To capture long-distance dependencies between tokens, we draw inspiration from message passing frameworks to model tokens as entity representations, enabling one-hop connections based on their relation representations (i.e., task-relevant correlations).
In typical message passing, the output embedding $E$ is formulated by the concatenation of learned entity representations of each token as follows:
\begin{equation}\label{eq:entity}
    E = \mathrm{CONCAT}(e_1, e_2, ..., e_i),
\end{equation}
where $e_i$ is derived from the entity representation of token embedding $x_i$.
Therefore, the correlation between tokens can be modeled by the similarity between their entity representations.
To seamlessly integrate these learned correlations into LLMs, \adapter{} takes the key-value (KV) embeddings, which are cached by attention heads during generation, as the entity representation of tokens.
In other words, \adapter{} aims to encode the correlation between entity representations into the attention feature of tokens.
Formally, given the KV embeddings $[K_i, V_i]$ of each token, the entity encoding of \adapter{} first maps the concatenated embeddings into the entity representation through linear projection, which can be given by:
\begin{equation}\label{eq:proj_entity}
    e_i = \mathrm{Proj}_{\mathrm{entity}}(\mathrm{CONCAT(K_i, V_i)}),
\end{equation}
following Eq.~\eqref{eq:entity} definitions,  $e_i \in \mathbb{R}^{d_k}$ denotes the entity representation of each token through a linear layer $\mathrm{Proj}_{\mathrm{entity}}(\cdot)$.
Subsequently, the entity representation of tokens will be updated by their relation representations via message passing, as indicated by the message block in Fig.~\ref{fig:method}.

\subsubsection{Implementation of Relation Encoding}
In \adapter{}, the relation encoding aims to model the task-relevant correlation between tokens into the representation space of attention features.
To be specific, we obtain the task feature (i.e., recommendation knowledge) through the linear projection from learnable parameters for fine-tuning \llmrec{}.
For example, these learnable parameters can be acquired by 
soft prompt embeddings~\cite{liao2024llara,kim2024large, li2023prompt} and the readout of model parameters~\cite{geng2022recommendation} or parameter-efficient adapters~\cite{ji2024genrec, qu2024tokenrec}, which are typical fine-tuning paradigms of \llmrec{} used in existing works.
It is worth noting that the task feature extracted by these learnable parameters of \llmrec{} should not be attributed to additional \adapter{} parameters.
In other words, our \adapter{} can be integrated with almost any existing \llmrec{} paradigms, leveraging their learnable parameters as the task feature to encode the relation representation of tokens.
Formally, given the task feature $z\in\mathbb{R}^{d_z}$ and each entity representation $e_i$, we utilized their similarity
(i.e., task-relevant correlation) to encode the relation representation.
Similar to the entity encoding, a linear projection layer $\mathrm{Proj}_{\mathrm{relation}}(\cdot)$ is employed to first map the task feature into the representation space of attention features $d_k$, then the task-relevant correlation of tokens is formulated by:
\begin{equation}\label{eq:proj_relation}
    c_i = \bm{\mathrm{sim}}(\mathrm{Proj}_{\mathrm{relation}}(z), e_i),
\end{equation}
where $c_i \in \mathbb{R}^1$ denotes the importance score of each token derived from $\bm{\mathrm{sim}}(\cdot)$, such as cosine similarity, indicating the task-relevant correlation (e.g., relevant side information) in recommendation tasks.
Subsequently, the matrix of relation representations can be modeled by broadcasting the task-relevant correlation between tokens, which calculates the dot product of $c_i$ and $c_j (i \neq j)$.
However, the large volume of noisy web information may inevitably cause a massive $O(N^2)$ computational cost of the relation matrix, where $N$ is determined by the number of tokens.
Therefore, we further take advantage of the top-$k$ ranking ($k \ll N$) of task-relevant correlations to prune the relation between entities, reducing the computation cost of the relation matrix from $O(N^2)$ to $O(k^2)$.
To be specific, all-zero values are assigned to relation representations that are outside of the top-$k$ ranking, which indicates no connectivity between entities with relatively low task-relevant correlations.
This contributes to eliminating noisy information from certain web tokens while highlighting others.
Formally, given each pair of entities $e_i, e_j (i \neq j)$, the relation representation calculates the dot product of their task-relevant correlations as follows:
\begin{equation}\label{eq:relation}
    r_{i,j}= 
    \begin{cases}
        \max(0, c_i \cdot {c_j}^T),  & \text{if} \ i,j \in \ \mathcal{T}_k \\
        0, & \text{otherwise}
    \end{cases}
    ,
\end{equation}
where $\mathcal{T}_k$ is the list of entity indices based on the top-$k$ ranking of task-relevant correlations.
For example, in a setting of $k=1$, the top-$1$ list is given by $\mathcal{T}_k = \{\mathrm{arg\,max}_{k \in [1, 2, ..., N]}(c_k)\}$.
It is worth noting that the matrix of relation representations that $[r_{1,1}, r_{1,2}, ..., r_{N,(N-1)}, r_{N,N}]$ shares the same dimension as the matrix of attention scores, while their computational methods at attention heads are different.
Specifically, different from the weighted sum in Eq.~\eqref{eq:attention} generated by vanilla attention heads, the relation representation enables one-hop message passing between long-distance entities in \adapter{} that alleviates long-distance dependencies in the weighted sum.

\subsubsection{Implementation of Message Passing}
Following the entity and relation encoding, the message passing updates entity representations by aggregating neighbor entities based on their relation representations.
The key insight of integrating message passing into \adapter{} lies in capturing long-distance dependencies between the entity representations of web and recommendation tokens.
For example, high-quality tokens over noisy web information can be modeled as one-hop neighbors.
In particular, we capture these high-quality tokens by encoding their task-relevant correlations into relation representations, which indicates the weight for aggregating neighbor entities.
Notably, instead of updating entity representations with equal weights, the learnable relation representations contribute to eliminating noisy information from certain web tokens while highlighting others.
Thereby, the updated entity representations of task-relevant tokens via message passing can exhibit strong similarity with each other, guiding \adapter{} to capture their task-relevant correlations by generating high attention scores between these tokens.

Formally, in message passing at transformer layer $l$, the entity representation of each token can be updated based on their message of neighborhood aggregation. In particular, the message aggregates information from all task-relevant entities, which can be defined by:
\begin{equation}\label{eq:message}
    m_i^{(l)} \gets \mathop{\sum_{j \in \mathcal{T}_k}} \frac{r_{i,j}}{\sqrt{\bm{\mathrm{deg}}(i)}\sqrt{\bm{\mathrm{deg}}(j)}} \cdot \mathrm{Proj}_{\mathrm{entity}}(e_j^{(l-1)}) + b,
\end{equation}
where $\bm{\mathrm{deg}}(\cdot)$ denotes the degree of entity for normalizing neighbor entities and $b$ is a bias vector.
Subsequently, the entity representation is updated by a update function $\mathcal{U}_l$, such as $\bm{\mathrm{ReLU}}(m_i^{(l)})$~\cite{cui2024prompt} and $\mathrm{CONCAT(e_i^{(l-1)}, m_i^{(l)})}$~\cite{gilmer2017neural} or $e_i^{(l-1)} +m_i^{(l)}$~\cite{sun2024towards} depending on task objectives.
However, the update function in conventional message passing is merely responsible for capturing the similarity between task-relevant tokens, while \adapter{} aims to model such similarity as the attention between these tokens.
In other words, the attention score of tokens can be calculated by the similarity with their messages, indicating the importance of tokens in recommendation tasks.
Therefore, to model attentions based on updated entity representations, \adapter{} scores the entity representation of tokens under their representation space of attention features as follows:
\begin{equation}
    \mathcal{H}_l(Q_i^{(l)}, K_i^{(l)}) =  \bm{\mathrm{softmax}}(Q_i^{(l)} \cdot {e_i^{(l)}[:|K_i|]}^T),
\end{equation}
where $|K_i|$ is the length of key embeddings and the entity is updated by:
\begin{equation}\label{eq:update}
     e_i^{(l)} \gets \mathcal{U}_l(e_i^{(l-1)}, m_i^{(l)}) = m_i^{(l)}.
\end{equation}
Notably, the dot product of $Q_i \cdot {e_i^{(l)}[:|K_i|]}^T$ in \adapter{} is analogous to attention scores in vanilla attention heads that $Q_i \cdot K_i^T$, while the term $e_i^{(l-1)}$ can be ignored in the update function that Eq.~\eqref{eq:update} due to self-attention mechanisms (i.e., the message includes each entity themselves).

\subsubsection{Training of \llmrec{} with \adapter{}}
During the training stage of \llmrec{}, our proposed \adapter{} can be seamlessly updated by the recommendation loss.
However, the capability of \adapter{} can be heavily dependent on the learned task feature of recommendations that $z$ in Eq.~\eqref{eq:proj_relation}, which is not innately embedded in \llmrec{} before training.
To this end, we introduce a gating factor $g_i$ to control the integration of \adapter{} with vanilla attention heads during the training of \llmrec{}.
Formally, the output of \adapter{} can be formulated as:
\begin{equation}\label{eq:output}
   \mathbf{head}_{MP}(x_i) \gets g_i \cdot \mathrm{Attention_{MP}}(Q_i^{(l)}, K_i^{(l)}, V_i^{(l)}) = g_i \cdot \mathcal{H}_l(Q_i^{(l)}, K_i^{(l)})e_i^{(l)}[|V_i|:],
\end{equation}
where $|V_i|$ is the length of value embeddings.
It is worth noting that a learnable gating factor is critical to the effective learning of \adapter{}, since noisy web information may cause disturbance, especially at the early training stage of \llmrec{}.
More specifically, the gating factor alleviates normalizing the concatenation of \adapter{} with vanilla attention heads in Eq.~\eqref{eq:mp-head}, which prevents affecting the pre-trained distribution of attentions.
Therefore, we can regard the original \llmrec{} as a frozen model parameterized by $\Phi_0$, while the training parameters $\Theta$ are modeled by the learnable task feature, such as parameter-efficient prompt tuning, and our proposed \adapter{}.
Formally, given the input tokens of web-augmented prompt $x$ and the label tokens of recommendation output $y$ (e.g., target item titles), the training loss of \llmrec{} with \adapter{} is defined by the negative log-likelihood as follows:
\begin{equation}\label{eq:loss}
    \mathcal{L}(x, y) = - \mathop{\sum_{t=1}^{|y|}}\mathrm{log}(\mathop{\mathrm{Pr}_{\Phi_0 + \Delta \Phi(\Theta)}}(y_t | x, y_{<t})).
\end{equation}
To be specific, the parameters $\Theta$ involves both \adapter{} modules, including the gating factor $g_i$ and two linear projection layers (i.e., $\mathrm{Proj}_{\mathrm{entity}}, \mathrm{Proj}_{\mathrm{relation}}$), and the learnable task feature $z$.
It is worth noting that this task feature is obtained by the learnable parameters for fine-tuning \llmrec{}, such as prompt embeddings~\cite{liao2024llara,kim2024large, li2023prompt} or trainable adapter parameters~\cite{ji2024genrec, geng2022recommendation, qu2024tokenrec} used in existing works.
In other words, \adapter{} is not responsible for learning the task feature, but leveraging it to capture task-relevant correlations between web and recommendation tokens.

%% file: 04experiment.tex
\section{EXPERIMENT}

\subsection{Experimental Settings}

\subsubsection{Datasets}
To evaluate the performance of \ours{} in real-world recommendation scenarios, we conducted experiments on four datasets derived from Amazon databases~\cite{ni2019justifying}: \textit{Beauty}, \textit{Toys}, \textit{Video Games}, and \textit{Movies and TV}.
These datasets record rich user-item interactions and textual metadata, including user reviews, item titles, and descriptions.
To facilitate an extensive analysis, each benchmark dataset has a varying distribution in terms of the number of users and items, where the basic statistics are presented in Table~\ref{tab:dataset}.
Following existing studies~\cite{kim2024large, liao2024llara}, we preprocess each dataset by removing users and items with fewer than five interactions.
To construct sequential recommendation scenarios, we adopt the leave-one-out strategy and retain the first 10 items in each sequence as the historical interaction, and the last item as the target item.
Note that interaction sequences with fewer than 10 interactions are padded.
For both datasets, we split the data points of user-item interactions into training, validation, and testing sets with a ratio of 8:1:1, which excludes subsequent interactions from the training dataset and prevents information leakage.

As for retrieval databases in RAG, we consider both local knowledge in recommendations (i.e., Amazon datasets) and external knowledge from online websites.
The basic statistics of retrieval tools are presented in Table~\ref{tab:web}.
In particular, Amazon datasets are formatted by metadata, where the unique identifier (UID) is available to retrieve user reviews (e.g., rating and comment) and item features, including price, description, and sales information.
It is worth noting that, to categorize user–item feedback for retrieval, ratings above 3 are regraded as positive reviews, whereas ratings of 3 or below, as well as unobserved items, are treated as negative reviews.
In terms of web search results, noisy retrieved information is presented in plain text, such as page title, page snippet (i.e., a short paragraph describing page content), and page link generated by search APIs.

\begin{table*}[h]
\begin{minipage}{0.48\linewidth}
    \centering
    \caption{The basic statistics of benchmark datasets, where $\overline{L}_{\mathrm{user}}$ denotes the average length of the interaction sequence of each user.}
    \label{tab:dataset}
    \resizebox{\linewidth}{!}{
    \begin{threeparttable}
    \begin{tabular}{l | ccc | l}
        \toprule
        \textbf{Datasets} & \#User & \#Item & \#Interaction & $\overline{L}_{\mathrm{user}}$\\
        \midrule
        Beauty & 9,930 & 6,141 & 63,953 & $\approx 6$ \\
        Toys & 30,831 & 61,081 & 282,213 & $\approx 9$ \\
        Video Games & 64,073 & 33,614 & 598,509 & $\approx 8$ \\
        Movies and TV & 297,498 & 59,944 & 3,409,147 & $\approx11$ \\
      \bottomrule
    \end{tabular}
    \begin{tablenotes}
        \item \# denotes the number of users, items, and interactions
    \end{tablenotes}
    \end{threeparttable}
    }
\end{minipage}
\hfill
\begin{minipage}{0.48\linewidth}
    \centering
    \caption{The basic statistics of retrieval tools. Tavily and Brave are two popular web search APIs, where each retrieval result contains: page title, page HTML, and page snippet (i.e., description paragraph).}
    \label{tab:web}
    \resizebox{\linewidth}{!}{
    \begin{threeparttable}
    \begin{tabular}{l | c | ccc}
        \toprule
        \textbf{Retrieval Tools} & Format & User review & Item feature & Website \\
        \midrule
        Amazon datasets$^1$ & metadata & \yessymbol (UID) & \yessymbol (UID) & \nosymbol \\
        Tavily search API$^2$ & plain text & \nosymbol & \nosymbol & \yessymbol (Noisy) \\
        Brave search API$^3$ & plain text & \nosymbol & \nosymbol & \yessymbol (Noisy) \\
      \bottomrule
    \end{tabular}
    \begin{tablenotes}
        \item $^1$ https://jmcauley.ucsd.edu/data/amazon/
        \item $^2$ https://www.tavily.com/ $^3$ https://www.brave.com/
    \end{tablenotes}
    \end{threeparttable}
    }
\end{minipage}
\end{table*}

\subsubsection{Baselines}
In the field of LLM-based recommendations, our work pioneers the research on exploring web-based RAG to facilitate the performance of \llmrec{}. 
Therefore, to evaluate the effectiveness of our proposed \ours{}, we compare the recommendation performance with the following baselines, which can be categorized into three types:
\begin{itemize}[leftmargin=*, noitemsep, topsep=0pt]
    \item \textbf{Non-RAG} for \llmrec{}. POD~\cite{li2023prompt} transforms discrete prompt representations into continuous embeddings in order to mitigate the issue of excessive input length in LLMs, building upon the P5 foundational models. GenRec~\cite{ji2024genrec} leverages the understanding capabilities of LLMs to generate personalized recommendations, which perform instruction tuning on recommendation datasets based on the LLaMA foundational models.
    Similarly, TALLRec~\cite{bao2023tallrec} emulates the pattern of instruction tuning to align LLMs with user preferences in recommendation tasks.
    
    \item \textbf{Recommendation-based RAG} for \llmrec{}. Most existing works on RAG-enhanced \llmrec{} are limited to matching-based methods, where user/item IDs are available for accurate retrieval.
    CoRAL~\cite{wu2024coral} aligns LLM reasoning with task-specific user–item interaction knowledge by retrieving collaborative evidence directly into prompts.
    RA-Rec~\cite{kemper2024retrieval} utilizes comprehensive item reviews retrieved from known metadata to facilitate closer alignment with users’ potential interests.
    ROPG~\cite{salemi2024optimization} attempts to optimize retrieval models that supply a constrained set of personal documents to LLMs, and enhance personalized recommendations.
    For these baselines, we employ the recommendation metadata as the RAG database to retrieve relevant information, including user reviews, product descriptions, and sales information (i.e., ``also buy'' interactions).
    
    \item \textbf{Web-based RAG} for \llmrec{}. Most existing works on web-based RAG are designed for general LLMs, such as open-domain QA, which fall short in addressing the knowledge gap of task-specific recommendations.
    Self-RAG~\cite{asai2024self} enhances the generation quality of LLMs by combining retrieval with self-reflection, where LLMs adaptively retrieve relevant passages and refine their own outputs.
    CRAG~\cite{yan2024corrective} employs LLMs as retrieval evaluators to inform and regulate the selection of alternative knowledge retrieval strategies, which selectively filter out irrelevant content in retrieved information.
    RAG drafter~\cite{wang2024speculative} proposes to evaluate multiple draft responses generated in parallel by specialist LLMs, where each draft is constructed from a distinct subset of retrieved information, producing complementary perspectives on the available evidence.
\end{itemize}

\subsubsection{Implementation Details}
We implement the proposed framework on the basis of HuggingFace and PyTorch, in which all the experiments are conducted on two NVIDIA H20-96GB GPUs.
For each search API of web retrieval, we preserve the top-$10$ websites as the default setting in RAG paradigms, and more comprehensive results for $k=1, 5, 10, 15, 20$ are present in ablation studies.
To implement our proposed \adapter{}, following common settings of message passing~\cite{wang2025knowledge, cui2024prompt}, we compare the performance of one to three head layers, which are inserted into the Transformer blocks of LLMs, including block layers of $0, 2, 15$ and $31$.
In particular, we employ LLaMA$_{7B}$ and LLaMA-2$_{7B}$ as the backbone model of LLMs, where the layer dimension is set to 4096, and the attention head number is set to 32.
In the training stage, we optimize the \ours{}  with AdamW while freezing the LLM backbone, where the training epoch is set to 5 with a batch size of 8.

\subsubsection{Evaluation Metrics}
To evaluate the effectiveness of our \ours{}, we employ two widely used evaluation metrics: the top-$k$ hit rate (\textbf{HR@$k$}) and the top-$k$ normalized discounted cumulative gain (\textbf{NDCG@$k$} or \textbf{NG@$k$} for simplicity), in which large values indicate higher recommendation performance.
We select $k=5, 10$ as the default value for the main experiments, and more comprehensive results for $k=1, 3, 5, 10, 20$ are present in ablation studies.
In pursuit of fair comparisons between our proposed \ours{} and existing \llmrec{} paradigms~\cite{kim2024large, liao2024llara}, for sequential recommendation scenarios, we augment the interaction sequence by randomly selecting 20 non-interacted items as the candidate set.
That is, the test set for each user contains one positive term (i.e., target item) and 19 negative terms (i.e., non-interacted items).
To adapt LLMs to recommendation tasks, these interaction sequences are converted into textual recommendation prompts.
In addition, the retrieved information based on \ours{} and other RAG baselines will be concatenated to recommendation prompts, where an example template is provided as follows:

\begin{tcolorbox}[
fonttitle=\fon{pbk}\bfseries,
fontupper=\sffamily,
fontlower=\sffamily,
enhanced,
left=2pt, right=2pt, top=2pt, bottom=2pt,
title=Prompt Template: Web-based RAG for Recommendations,
width=0.95\linewidth,
center
]
\textbf{[INPUT]}

\textbf{\textit{(Instruction)}}
Below is an instruction that describes a task, paired with webpages that provide further context. Write a response that appropriately completes the request.

\textbf{\textit{(Recommendation Prompt)}}
A user has bought $\langle \text{item}\_1 \rangle$, $\langle \text{item}\_2 \rangle$, ..., and $\langle \text{item}\_10 \rangle$ in the previous. Please recommend the next item for this user to buy from the following item title set: $\langle \text{candidate}\_1 \rangle$, $\langle \text{candidate}\_2 \rangle$, ..., and $\langle \text{candidate}\_20 \rangle$. The recommendation is

\textbf{\textit{(Websites)}}
[1] Title: $\langle \text{page}\_\text{title} \rangle$. Content: $\langle \text{page}\_\text{snippet} \rangle$;
[2] Title: $\langle \text{page}\_\text{title} \rangle$. Content: $\langle \text{page}\_\text{snippet} \rangle$;
... ...;
[10] Title: $\langle \text{page}\_\text{title} \rangle$. Content: $\langle \text{page}\_\text{snippet} \rangle$.

\tcblower

\textbf{[OUTPUT]}

\textbf{\textit{(Target Item)}}
$\langle \text{candidate}\_\text{n} \rangle$
\end{tcolorbox}

\subsection{Performance Comparison}
As shown in Table~\ref{tab:big-table}, our proposed method demonstrates state-of-the-art performance across all four recommendation domains, demonstrating significant improvements over existing baselines.
When integrated with Tavily API, our method yields relative gains from $8.7\%$ to $19.2\%$ in HR metrics and $4.5\%$ to $17.2\%$ in NDCG metrics over the strongest baselines.
Similarly, with Brave API, improvements range from $5.2\%$ to $23.0\%$ in HR metrics and $1.4\%$ to $13.9\%$ in NDCG metrics.
This consistent superiority underscores our framework's ability (i.e., \adapter{}) to effectively capture relevant recommendation information over noisy web data.
Building upon these findings, we elaborate on detailed observations in the following sections.

\subsubsection{Limitations of Non-RAG Methods}
Non-RAG methods exhibit noticeable constraints in recommendation quality due to their reliance on limited parametric knowledge of pre-trained LLMs.
For example, though TALLRec that integrates direct user preference into prompts can achieve competitive results compared to recommendation-based RAG baselines, the performance degrades significantly by $24.9\%$ from our best performance.
Notably, such a performance gap widens further for POD and GenRec baselines, which merely take user-item interactions.
These limitations prove inadequate for modeling complex user preferences without external knowledge augmentation.

\begin{table*}[t]
\renewcommand{\arraystretch}{1.2} 
\setlength\tabcolsep{3pt}
\caption{
Performance comparison between our proposed \ours{} and baselines, where \textbf{boldface} and \underline{underline} indicate the best and second best scores for each column, respectively. We independently report the results of web-based RAG under different sources of web data, including \colorbox{atomictangerine!20}{Taily} and \colorbox{ballblue!15}{Brave} search APIs. Our improvement is compared with the best baseline (i.e., excluding ours).  
}
\label{tab:big-table}

\resizebox{\linewidth}{!}{
\begin{threeparttable}
\begin{tabular}{l | cccc cccc cccc cccc}
\toprule[1pt]

& \multicolumn{4}{c}{\textbf{Beauty}} 
& \multicolumn{4}{c}{\textbf{Toys}} 
& \multicolumn{4}{c}{\textbf{Video Games}} 
& \multicolumn{4}{c}{\textbf{Movies and TV}} \\

\cmidrule(lr){2-5} \cmidrule(lr){6-9} \cmidrule(lr){10-13} \cmidrule(lr){14-17} 

& HR@5 & HR@10 & NG@5 & NG@10
& HR@5 & HR@10 & NG@5 & NG@10
& HR@5 & HR@10 & NG@5 & NG@10
& HR@5 & HR@10 & NG@5 & NG@10 \\

\midrule\midrule

\multicolumn{17}{l}{\textbf{Non-RAG (\textit{w/o} retrieval, i.e., recommendation prompt only)}} \\

POD~\cite{li2023prompt}
    & 0.3166 & 0.3354 & 0.2864 & 0.2928
    & 0.5729 & 0.5806 & 0.5448 & 0.5473
    & 0.4692 & 0.5428 & 0.3783 & 0.4105
    & 0.5001 & 0.5714 & 0.4221 & 0.4532 \\
GenRec~\cite{ji2024genrec}
    & 0.3345 & 0.3475 & 0.2909 & 0.3038
    & 0.5559 & 0.5625 & 0.5424 & 0.5446
    & 0.4705 & 0.5419 & 0.3713 & 0.3962
    & 0.5463 & 0.5717 & 0.4456 & 0.4581 \\
TALLRec~\cite{bao2023tallrec}
    & 0.3533 & 0.3644 & 0.2978 & 0.3109
    & 0.5848 & 0.6096 & 0.5465 & 0.5616
    & 0.4725 & 0.5440 & 0.3899 & 0.4188
    & 0.5602 & 0.5798 & 0.4508 & 0.4592 \\

\midrule\midrule

\multicolumn{17}{l}{\textbf{Recommendation-based RAG (\textit{w/} retrieval from Amazon metadata$^1$, e.g., item reviews)}} \\

RA-Rec~\cite{kemper2024retrieval}
    & 0.3114 & 0.3727 & 0.2775 & 0.3056 
    & 0.5511 & 0.6077 & 0.5425 & 0.5642 
    & 0.5269 & 0.5632 & 0.4064 & 0.4210 
    & 0.5226 & 0.5898 & 0.4398 & 0.4730 \\
ROPG~\cite{salemi2024optimization}
    & 0.2748 & 0.3530 & 0.2677 & 0.3034
    & 0.5438 & 0.5926 & 0.5318 & 0.5554 
    & 0.5286 & 0.5570 & 0.4045 & 0.4172 
    & 0.5228 & 0.5862 & 0.4315 & 0.4662 \\
CoRAL~\cite{wu2024coral}
    & 0.3655 & 0.3878 & 0.3002 & 0.3102 
    & 0.5997 & 0.6165 & 0.5606 & 0.5690 
    & 0.5015 & 0.5754 & 0.3931 & 0.4239 
    & 0.5670 & 0.6019 & 0.4680 & 0.4814 \\
    
\midrule\midrule

\rowcolor{atomictangerine!20}  \multicolumn{17}{l}{\textbf{Web-based RAG (\textit{w/} retrieval from Tavily search API$^2$)}} \\

Self-RAG~\cite{asai2024self}
    & 0.3298 & 0.3319 & 0.2849 & 0.2882
    & 0.5328 & 0.5636 & 0.5308 & 0.5505
    & 0.4686 & 0.5522 & 0.3632 & 0.3949
    & 0.5517 & 0.5698 & 0.4377 & 0.4407 \\
CRAG~\cite{yan2024corrective}
    & 0.3392 & 0.3400 & 0.2919 & 0.2921
    & 0.5402 & 0.5453 & 0.5372 & 0.5403
    & 0.4736 & 0.5417 & 0.3672 & 0.3908
    & 0.5474 & 0.5635 & 0.4403 & 0.4501 \\
RAG drafter~\cite{wang2024speculative}
    & 0.3716 & \underline{0.4319} & 0.3016 & 0.3187
    & 0.5897 & 0.5988 & 0.5606 & 0.5627
    & 0.4931 & 0.5692 & 0.3847 & 0.4161
    & 0.5606 & 0.5921 & 0.4612 & 0.4874 \\

\midrule

\ours{} (\textit{Ours})
    & \textbf{0.4043} & \textbf{0.4921} & \textbf{0.3153} & \textbf{0.3441}
    & \underline{0.6655} & \underline{0.7352} & \underline{0.5871} & \underline{0.6098}
    & \textbf{0.5695} & \textbf{0.6565} & \textbf{0.4688} & \textbf{0.4970}
    & \underline{0.5880} & \textbf{0.7056} & \underline{0.4950} & \textbf{0.5346} \\

\textbf{\textit{Improvement}}
    & \textbf{8.79\%} & \textbf{13.93\%} & \textbf{4.54\%} & \textbf{7.96\%}
    & \textbf{10.97\%} & \textbf{19.25\%} & \textbf{4.72\%} & \textbf{7.17\%}
    & \textbf{13.55\%} & \textbf{14.09\%} & \textbf{15.35\%} & \textbf{17.24\%}
    & \textbf{3.70\%} & \textbf{17.22\%} & \textbf{5.76\%} & \textbf{9.68\%} \\

\midrule\midrule

\rowcolor{ballblue!15}  \multicolumn{17}{l}{\textbf{Web-based RAG (\textit{w/} retrieval from Brave search API$^3$)}} \\

Self-RAG~\cite{asai2024self}
    & 0.3218 & 0.3251 & 0.2891 & 0.2907
    & 0.5185 & 0.5223 & 0.5125 & 0.5205
    & 0.4799 & 0.5322 & 0.3646 & 0.3862
    & 0.5427 & 0.5683 & 0.4439 & 0.4461 \\
CRAG~\cite{yan2024corrective}
    & 0.3347 & 0.3355 & 0.2912 & 0.2953
    & 0.5358 & 0.5390 & 0.5318 & 0.5382
    & 0.4735 & 0.5365 & 0.3667 & 0.3971
    & 0.5470 & 0.5614 & 0.4434 & 0.4527 \\
RAG drafter~\cite{wang2024speculative}
    & 0.3783 & 0.4076 & 0.3024 & 0.3195
    & 0.5994 & 0.6043 & 0.5566 & 0.5635
    & 0.5075 & 0.5834 & 0.3874 & 0.4172
    & 0.5619 & 0.6081 & 0.4632 & 0.4852 \\

\midrule

\ours{} (\textit{Ours})
    & \underline{0.3981} & \textbf{0.4921} & \underline{0.3068} & \underline{0.3373}
    & \textbf{0.6862} & \textbf{0.7614} & \textbf{0.5950} & \textbf{0.6192}
    & \underline{0.5520} & \underline{0.6384} & \underline{0.4450} & \underline{0.4830}
    & \textbf{0.6025} & \underline{0.6721} & \textbf{0.5106} & \underline{0.5330} \\

\textbf{\textit{Improvement}}
    & \textbf{5.23\%} & \textbf{20.07\%} & \textbf{1.45\%} & \textbf{5.57\%}
    & \textbf{14.42\%} & \textbf{23.05\%} & \textbf{6.13\%} & \textbf{8.82\%}
    & \textbf{4.42\%} & \textbf{9.42\%} & \textbf{9.49\%} & \textbf{13.94\%}
    & \textbf{6.26\%} & \textbf{10.50\%} & \textbf{9.10\%} & \textbf{9.85\%}  \\
    
\bottomrule[1pt]
\end{tabular}
\begin{tablenotes}
    \item $^1$ https://jmcauley.ucsd.edu/data/amazon/ $^2$ https://www.tavily.com/ $^3$ https://www.brave.com/
\end{tablenotes}
\end{threeparttable}
}
\end{table*}

\subsubsection{RAG with Noisy Web Information}
Compared to recommendation-based RAG, it can be observed that web-based RAG baselines barely contribute to facilitating recommendation performance, failing to harness noisy web information.
For instance, under Tavily API, RAG-Drafter improves only marginally over recommendation-based CoRAL up to $1.7\%$ in Amazon Beauty, while degrading performance in other domains up to $2.9\%$.
More critically, Self-RAG and CRAG consistently underperform recommendation-based RAG across metrics, revealing that naive web retrieval lacks effective mechanisms to bridge the significant gap between web and recommendations.

\subsubsection{Web Information Sources}
Our method establishes superior performance consistency across both search APIs, demonstrating remarkable robustness to information source variations.
To be specific, with Tavily API, we achieve relative gains from $13.9\%$ to $19.2\%$ over the strongest baselines.
Remarkably, this superiority extends to Brave API, where our method maintains dominant performance with improvements from $5.2\%$ to $23.0\%$ across all evaluation metrics.
In conclusion, our framework excels in leveraging web content for diverse recommendation domains, validating the ability to capture task-relevant information over noisy web data from different sources.

\subsection{Ablation Study}
To evaluate the effectiveness of the proposed key components, we conducted ablation experiments on the Amazon \textit{Beauty} and \textit{Toys} datasets.
In particular, we compare the distinctive influence of our proposed methods across the retrieval stage and the generation stage of RAG pipelines.
Overall, it can be observed that removing any component of the framework consistently degrades recommendation performance, which demonstrates the effectiveness and necessity of each module.
The detailed analysis of each ablation variant is illustrated in the following sections.

\subsubsection{Number of \adapter{} Layers.}
To further understand the role of \adapter{} in capturing long-distance dependencies through message passing (i.e., $n$-hop neighbor aggregation), we compare the recommendation performance with one to three \adapter{} layers, as well as the ablation without \adapter{} in Table~\ref{tab:layers}.
The results demonstrate that the 1-hop setting, analogous to the dimension of attention metrics, provides baseline improvements up to $25.2\%$ over no aggregation.
Meanwhile, increasing hop count contributes to enhancing performance by capturing deeper task-relevant correlation between tokens, as shown by 3-hop superiority in Amazon Beauty, surpassing 2-hop and 1-hop settings up to $8.4\%$ and $37.8\%$, respectively.
This suggests multi-hop aggregation captures complex dependencies beyond vanilla attentions.
However, excessive hops can cause degradation due to over-aggregation, evident in Amazon Toys, where the 3-hop setting consistently underperforms the 2-hop setting across all evaluation metrics. In other words, including distant neighbors can lead to a reduction in recommendation performance, which is likely due to over-smoothing task-relevant features.

\begin{table}[tbp]
\centering
    \begin{minipage}{0.48\textwidth}
    \centering
    \caption{Ablation on the number of \adapter{} layers (i.e., $n$-hop neighbor aggregation). The improvement compares the \textbf{\textcolor{ForestGreen}{best}} and \textbf{\textcolor{Maroon}{worst}} results.}
    \label{tab:layers}
    \resizebox{0.99\textwidth}{!}{ %
    \begin{tabular}{l |  c c c c}
    \toprule
     \textbf{Component} & \textbf{HR@5} & \textbf{HR@10} & \textbf{NG@5} & \textbf{NG@10} \\
     
    \midrule\midrule
    
    \rowcolor{atomictangerine!20} \multicolumn{5}{l}{
        \textbf{\adapter{}} (Amazon Beauty, Tavily search API)} \\
    
    - w/ 1-hop
        & 0.4043 & 0.4921 & 0.3153 & 0.3441 \\
    - w/ 2-hop
        & 0.4106 & 0.4796 & 0.3223 & 0.3444 \\
    - w/ 3-hop
        & \bf \color{ForestGreen}{0.4451} 
        & \bf \color{ForestGreen}{0.5203} 
        & \bf \color{ForestGreen}{0.3479} 
        & \bf \color{ForestGreen}{0.3721} \\
    - w/o \adapter{}
        & \bf \color{Maroon}{0.3228} 
        & \bf \color{Maroon}{0.4231} 
        & \bf \color{Maroon}{0.2533} 
        & \bf \color{Maroon}{0.2856} \\
        
    \midrule
    
    \textbf{\textit{Improvement}}
        & \textbf{37.88\%} & \textbf{22.97\%} & \textbf{37.34\%} & \textbf{30.28\%} \\
        
    \midrule\midrule
    
    \rowcolor{atomictangerine!20} \multicolumn{5}{l}{
        \textbf{\adapter{}} (Amazon Toys, Tavily search API)} \\
        
    - w/ 1-hop
        & 0.6655 & 0.7352 & 0.5871 & 0.6098 \\
    - w/ 2-hop
        & \bf \color{ForestGreen}{0.7037} 
        & \bf \color{ForestGreen}{0.7832} 
        & \bf \color{ForestGreen}{0.6238} 
        & \bf \color{ForestGreen}{0.6496} \\
    - w/ 3-hop
        & 0.7026 & 0.7690 & 0.6184 & 0.6397 \\
    - w/o \adapter{}
        & \bf \color{Maroon}{0.6176} 
        & \bf \color{Maroon}{0.7102} 
        & \bf \color{Maroon}{0.5122} 
        & \bf \color{Maroon}{0.5423} \\

    \midrule
    
    \textbf{\textit{Improvement}}
        & \textbf{13.94\%} & \textbf{10.27\%} & \textbf{21.78\%} & \textbf{19.78\%} \\

    \bottomrule
    \end{tabular}%
    }
\end{minipage}
\hfill
\begin{minipage}{0.48\textwidth}
    \centering
    \caption{Ablation on the position of \adapter{} layer (i.e., Transformer blocks of 32-layer LLMs). The improvement compares the \textbf{\textcolor{ForestGreen}{best}} and \textbf{\textcolor{Maroon}{worst}} results.}
    \label{tab:position}
    \resizebox{0.98\textwidth}{!}{ %
    \begin{tabular}{l |  c c c c}
    \toprule
    \textbf{Component} & \textbf{HR@5} & \textbf{HR@10} & \textbf{NG@5} & \textbf{NG@10} \\
     
    \midrule\midrule
    
    \rowcolor{atomictangerine!20} \multicolumn{5}{l}{
        \textbf{\adapter{}} (Amazon Beauty, Tavily search API)} \\
    - layer 0
        & 0.3907 & 0.4727 & 0.2985 & 0.3196 \\
    - layer 2
        & \bf \color{ForestGreen}{0.4043} 
        & \bf \color{ForestGreen}{0.4921} 
        & \bf \color{ForestGreen}{0.3153} 
        & \bf \color{ForestGreen}{0.3441} \\
    - layer 15
        & 0.3605 & 0.4545 & 0.2780 & 0.3082 \\
    - layer 31
        & \bf \color{Maroon}{0.3542} 
        & \bf \color{Maroon}{0.4514} 
        & \bf \color{Maroon}{0.2599} 
        & \bf \color{Maroon}{0.2904} \\

    \midrule

    \textbf{\textit{Improvement}}
    & \textbf{14.14\%} & \textbf{9.01\%} & \textbf{20.62\%} & \textbf{18.49\%} \\
    
    \midrule\midrule
    
    \rowcolor{atomictangerine!20} \multicolumn{5}{l}{
        \textbf{\adapter{}} (Amazon Toys, Tavily search API)} \\
    - layer 0
        & 0.6498 & 0.7243 & 0.5625 & 0.5744 \\
    - layer 2
        & \bf \color{ForestGreen}{0.6655} 
        & \bf \color{ForestGreen}{0.7352} 
        & \bf \color{ForestGreen}{0.5871} 
        & \bf \color{ForestGreen}{0.6098} \\
    - layer 15
        & 0.6597 & 0.7305 & 0.5702 & 0.5933 \\
    - layer 31
        & \bf \color{Maroon}{0.6184} 
        & \bf \color{Maroon}{0.7135} 
        & \bf \color{Maroon}{0.5161} 
        & \bf \color{Maroon}{0.5512} \\

    \midrule

    \textbf{\textit{Improvement}}
    & \textbf{7.60\%} & \textbf{3.04\%} & \textbf{13.75\%} & \textbf{10.63\%} \\
    
    \bottomrule
    \end{tabular}%
    }
\end{minipage}
\end{table}

\subsubsection{Position of \adapter{} Layer.}
Following the number of \adapter{} layers, we further investigate the influence of layer positions as shown in Table~\ref{tab:position}.
The results reveal that the position of \adapter{} layers can significantly impact performance, where earlier layers generally yield higher performance across all metrics.
In particular, the insertion at layer 2 consistently achieves the best metrics across both datasets (i.e., improvement up to $20.6\%$), while layer 31 performs the worst.
We attribute this advantage to the task-specific correlations captured by front layers, which allow subsequent layers to effectively utilize these refined features.
Meanwhile, it is worth noting that inserting at the input layer (i.e., layer 0) underperforms relative to layer 2 by $1.4\%$ to $7.1\%$, despite being a front layer. 
We hypothesize that token representations remain weak during early attention layers of LLMs, limiting the ability of \adapter{} to generate accurate task-relevant correlations.

\subsubsection{Retrieval Strategy}
To evaluate the effectiveness of our retrieval strategy, we additionally compare the performance between the recommendation query (i.e., recommendation prompt) and the LLM-generated query that directly prompts LLMs to synthesize a web search query, such as ``\textit{Please generate the web search query for recommendations.}''.
As shown in Fig.~\ref{fig:ablation_retrieval}, web retrieval that directly uses recommendation queries shows marginal (i.e., up to $7.9\%$ in Amazon Beauty) or even negative (i.e., up to $2.3\%$ in Amazon Toys) gains versus the non-retrieval baseline, indicating a knowledge gap between web retrieval and recommendation tasks. 
It can be observed that our keyword-based strategy consistently outperforms both alternatives.
To be specific, compared to the non-retrieval baseline and LLM-generated query, our method achieves improvements up to $26.6\%$ and $7.4\%$, respectively, across all metrics in Amazon Toys.
Notably, these advantages retain or scale up to $46.7\%$ and $3.2\%$, respectively, in Amazon Beauty.
It is worth noting that our retrieval method can directly work on recommendation prompts, alleviating the necessity of fine-tuning LLMs or massive prompt engineering for retrieval tasks.

\subsubsection{Generation Strategy}
\begin{figure}[t] 
\centering    
    \begin{minipage}{0.49\textwidth}
    \centering
    \includegraphics[width=0.99\textwidth]{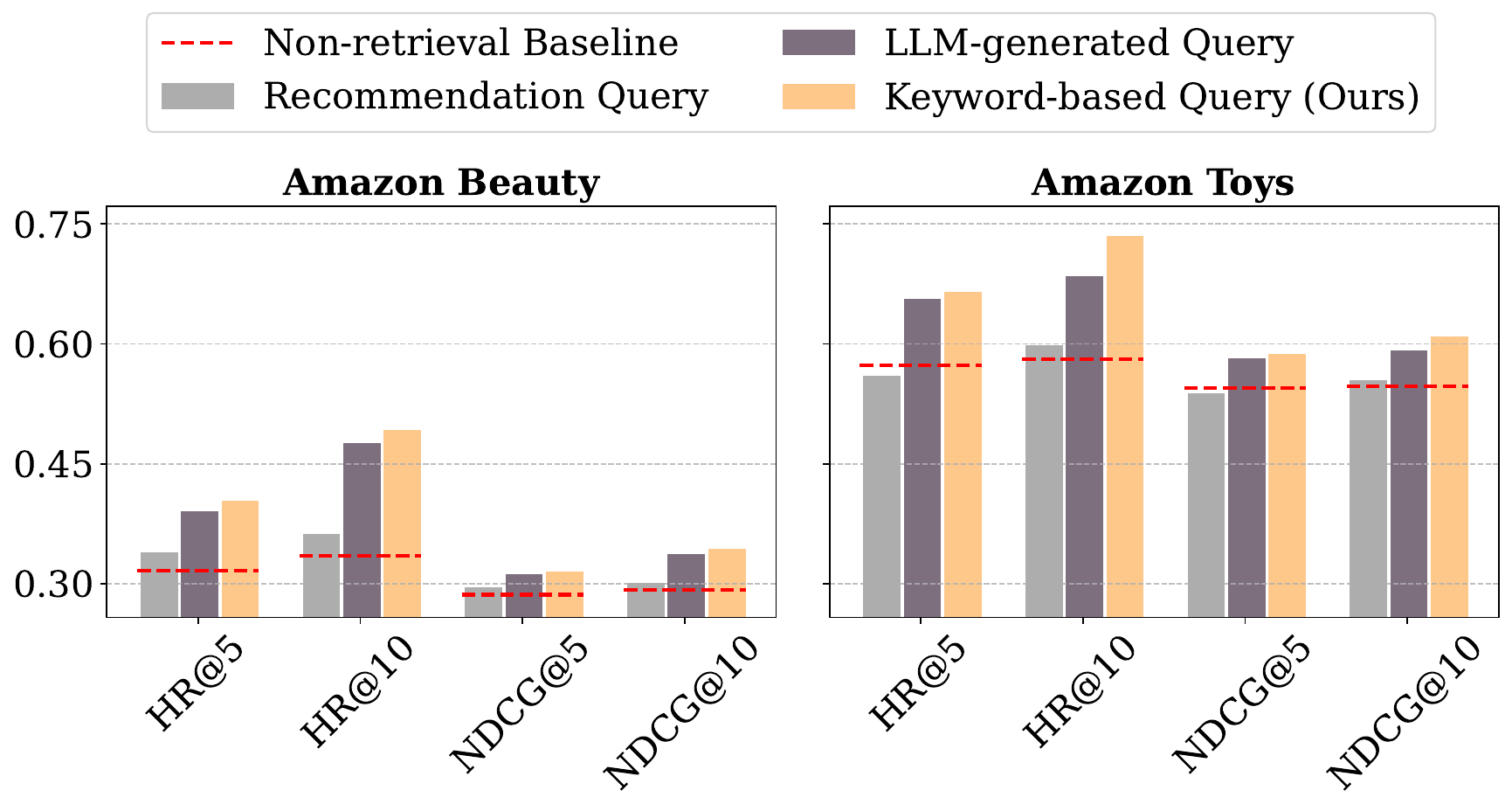}
    \caption{Ablation on Different Retrieval Strategies.}
    \label{fig:ablation_retrieval}
    \end{minipage}
    \hfill
    \begin{minipage}{0.49\textwidth}
    \centering
    \includegraphics[width=0.99\textwidth]{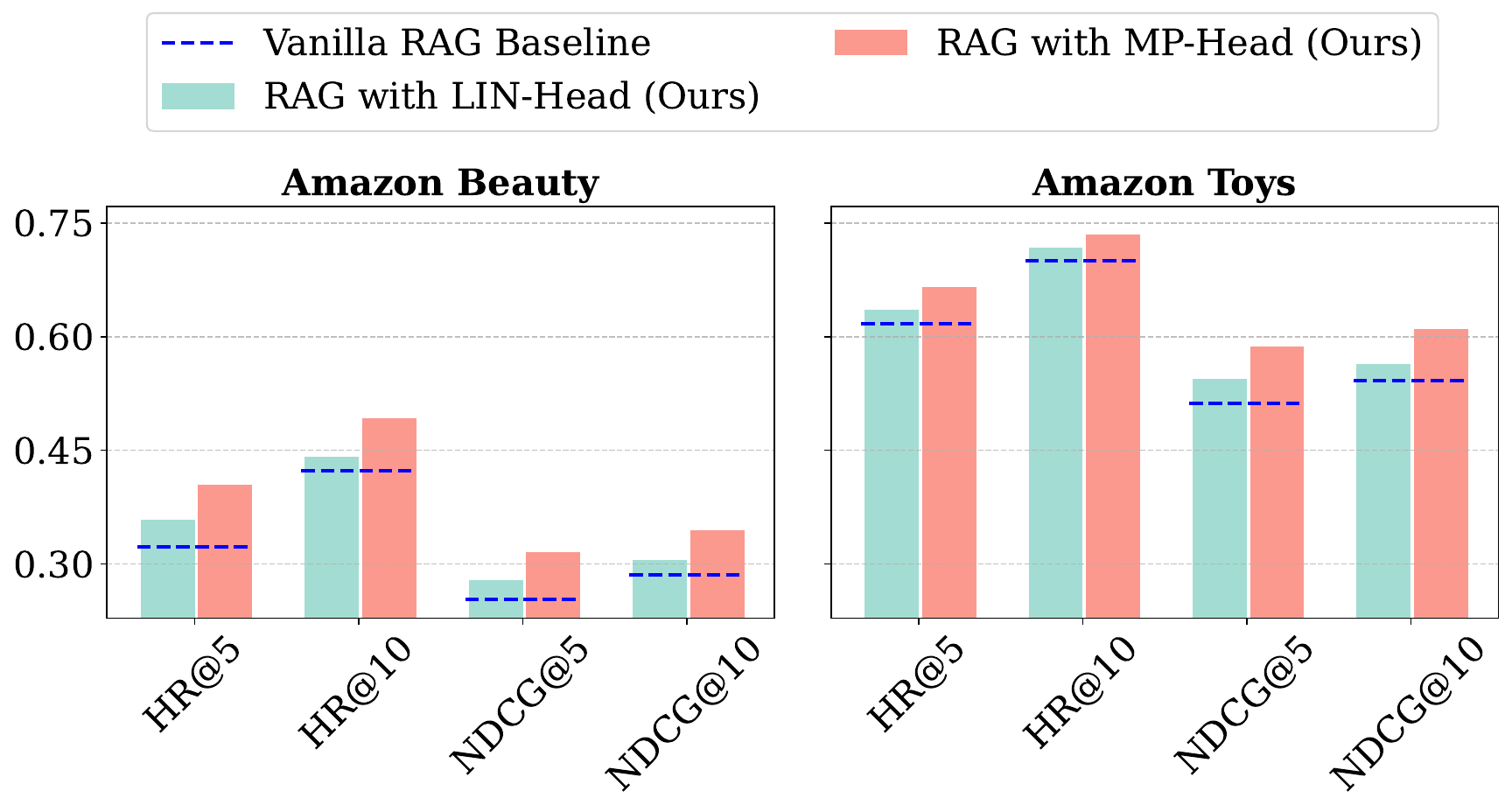}
    \caption{Ablation on Different Generation Strategies.}
    \label{fig:ablation_generation}
    \end{minipage}
\end{figure}

Due to the knowledge gap between web and recommendations, the retrieved side information can be noisy (e.g., large volume of irrelevant data), exhibiting long-distance semantic dependencies in recommendation tasks.
To assess the impact of our proposed \adapter{} on capturing task-relevant correlation over long-distance dependencies, we additionally compare the performance between vanilla RAG, where website content is naively concatenated as plain text, and our proposed variants, including LIN-head and \adapter{}.
In particular, LIN-head replaces the message passing structure by linear layers with equivalent dimensions, which is analogous to adding an attention head.
As illustrated in Fig.~\ref{fig:ablation_generation}, our method significantly outperforms the vanilla RAG baseline, demonstrating superior capability in harnessing noisy web information for recommendations.
On the Beauty dataset, MP-head achieves an absolute gain from $16.3\%$ to $25.2\%$ in HR metrics and from $20.4\%$ to $24.4\%$ in NDCG metrics.
Similar improvements are observed for Toys, where MP-head exceeds vanilla RAG up to $14.6\%$ across all metrics.
Notably, compared to LIN-head, our proposed message passing structure exceeds linear ones up to $13.0\%$ in both HR and NDCG metrics over Amazon Beauty. 
The above gains extend to Amazon Toys, with NDCG metrics showing the largest margin of $8.1\%$.
These findings underscore the effectiveness of our proposed \adapter{} in capturing task-relevant correlations between tokens by modeling their long-distance dependencies, compared to mere semantic correlations computed by linear attention heads.

\begin{figure}[t]
\centering
\subfigure[LLM Backbones]{
    \includegraphics[width=0.31\linewidth]{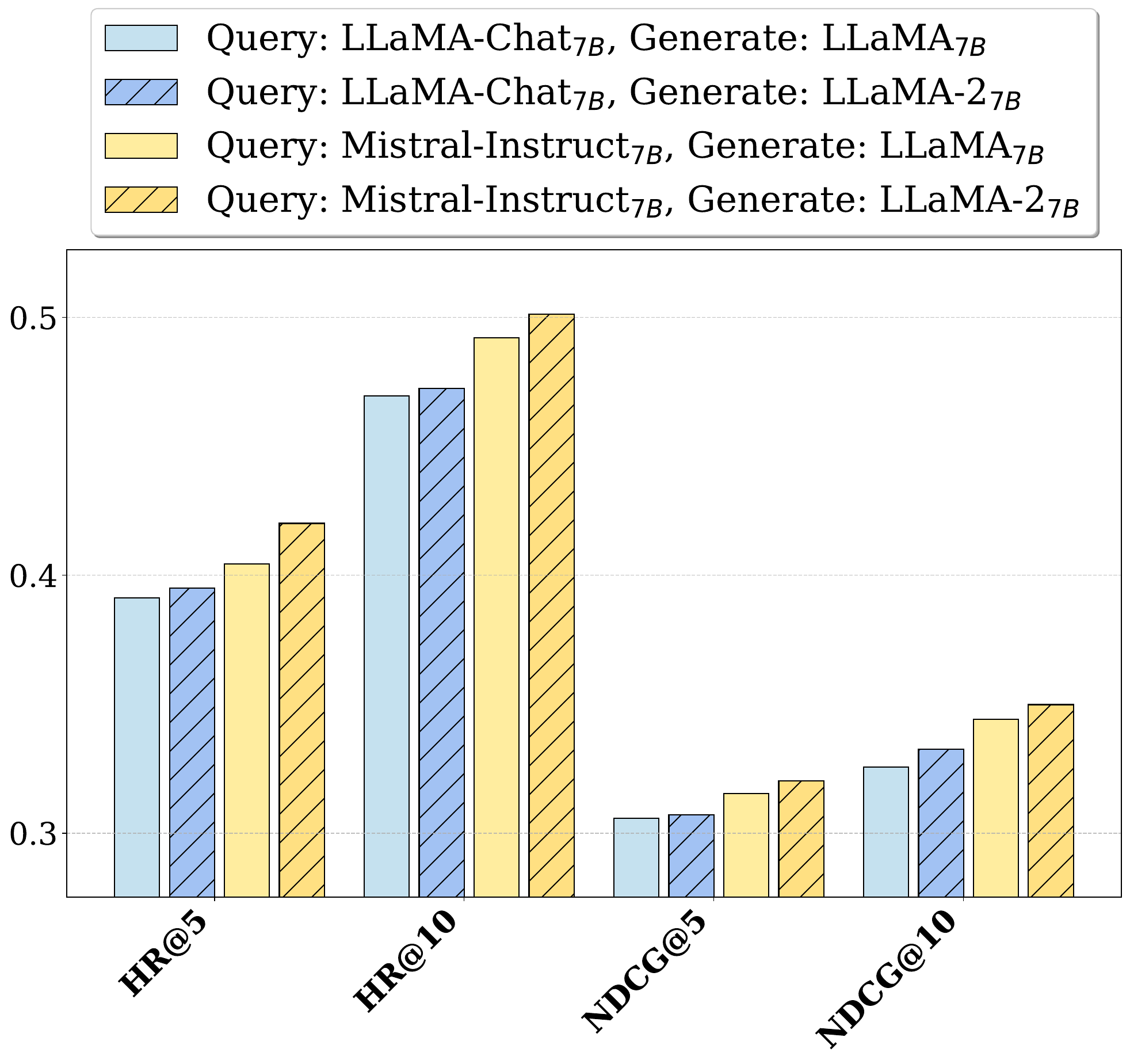}
    \label{fig:backbone}
}
\subfigure[Top-N Retrieval]{
    \includegraphics[width=0.31\linewidth]{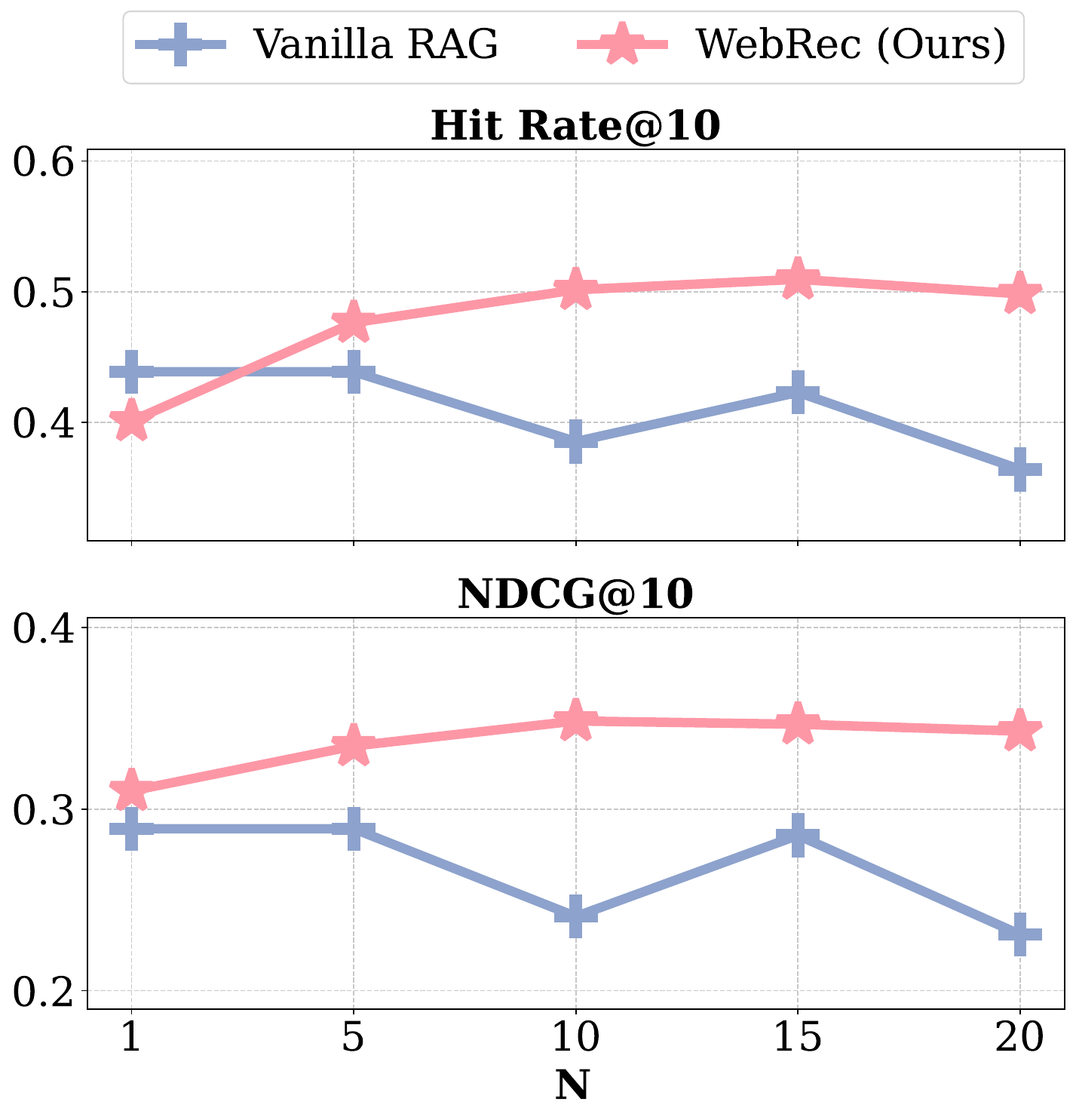}
    \label{fig:topn}
} 
 \subfigure[Top-K Recommendations]{
    \includegraphics[width=0.31\linewidth]{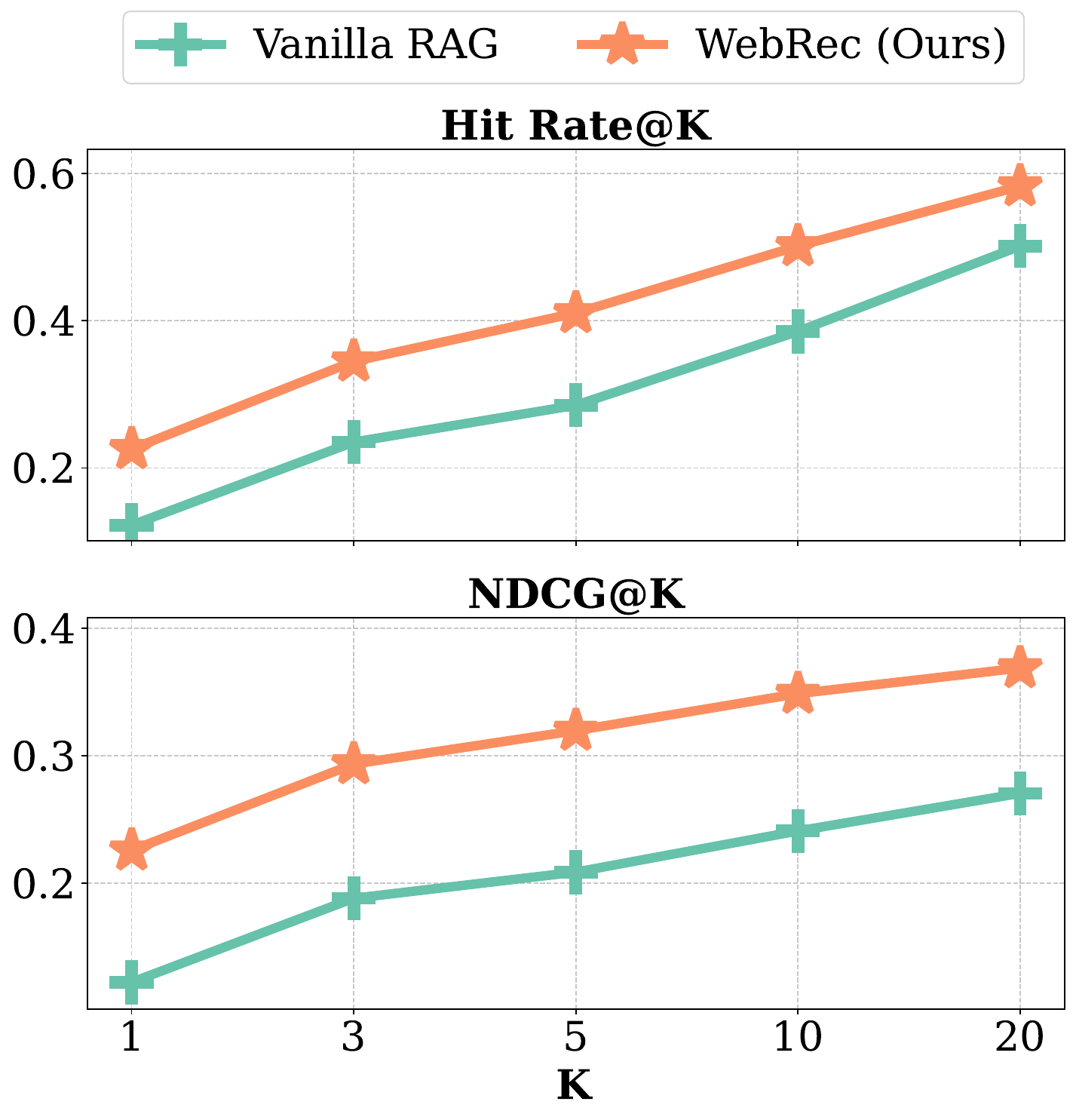}
    \label{fig:topk}
}
\caption{
Ablation on hyper-parameters of \ours{}. We report the results under Amazon Beauty.
}
\label{fig:ablation_hyperparameter}
\end{figure}

\begin{table*}[t]
\centering
\caption{Example of recommendation tasks and LLM-generated reasoning on user preferences.}
\scalebox{0.9}{
\begin{tabularx}{\textwidth}{X}
    \toprule
    \textbf{Recommendation Tasks (Prompt):} \\

    \midrule

    \textbf{\textcolor{purple}{A user has bought}} ``Crabtree amp; Evelyn - Gardener's Ultra-Moisturising Hand Therapy Pump - 250g/8.8 OZ'', ``Crabtree amp; Evelyn Gardeners Ultra-Moisturising Hand Cream Therapy - 3.5 oz'', ``Mustela Gentle Shampoo, Tear Free Baby Shampoo with Natural Avocado Perseose, Gently Cleanses and Detangles Kids' Hair, Available in 6.76 and 16.9 fl. oz'', and ``Noodle amp; Boo Super Soft Moisturizing Lotion for Daily Baby Care, Sensitive Skin and Hypoallergenic'' \textbf{\textcolor{purple}{in the previous. Recommend one next item for this user to buy next from the following item title set:}} ``GLYTONE KP Kit'', ``VINCENT LONGO Thinstick Lipstick'', ``Bioderma Atoderm Cream for Very Dry or Sensitive Skin'', ``La Roche-Posay Anthelios Ultra Light Sunscreen Fluid SPF 60, 1.7 Fl. Oz.'', ``Bioelements SPF 50 Face Screen, 2.3-Ounce'', ``Eau Thermale Avegrave;ne Antirougeurs Day Redness Relief Soothing SPF 25 Cream, 1.35 fl. oz.'', `` AHAVA Dead Mineral Botanic Velvet Cream Body Washes'', ``blinc Mascara'', ``Korres Body Butter, 7.95-Ounce Tube'', ``Zwilling J.A. Henckels Nose and Ear Hair Trimmer'', `` GLYTONE Exfoliating Body Lotion, 8.4 fl. oz.'', ``MERKUR Classic 2-Piece Double Edge Saftey Razor'', ``Eau Thermale Avegrave;ne Extremely Gentle Cleanser Lotion, 6.76 fl. oz.'', ``Glytone Daily Body Lotion Broad Spectrum Sunscreen SPF 15, 12 Ounce'', ``Mustela Stelatopia Cleansing Cream, Baby Body Wash for Extremely Dry to Eczema-Prone Skin, Fragrance-Free, 6.76 Fl. Oz'', ``HOT TOOLS Spring Curling Iron, Black/White'', ``St. Tropez Self Tan Bronzing Mousse 4 Fl Oz'', ``Vichy LiftActiv Retinol HA Eyes Anti-Aging Retinol Eye Cream with Hyaluronic Acid, 0.5 Fl. Oz.'', ``Philip B. Light-Weight Deep Conditioning Creme Rinse'', and ``MOROCCANOIL Intense Hydrating Mask Fragrance Originale, 16.9 Fl. Oz.''. \textbf{\textcolor{purple}{The recommendation is}} \\
    
    \midrule
    
    \textbf{Reasoning on User Preferences (LLM generation):} \\

    \midrule
    The user has bought hand therapy products and baby care products. Based on the user's previous purchases, it appears that they are interested in moisturizing and nourishing products for both their own hands and their baby's skin. Therefore, a recommendation for a body lotion or cream that is specifically formulated for dry or sensitive skin could be a good fit. \\
    One option from the given list that matches this description is the ``Bioderma Atoderm Cream for Very Dry or Sensitive Skin''. This product is known for its gentle and soothing formula, making it suitable for both adults and babies with dry or sensitive skin. It is also a cream, which is similar to the hand therapy and baby lotions the user has already purchased. \\
    Another option could be the ``Mustela Stelatopia Cleansing Cream, Baby Body Wash for Extremely Dry to Eczema-Prone Skin, Fragrance-Free, 6.76 Fl. Oz.'' This product is a baby body wash, but it is specifically formulated for very dry or eczema-prone skin, which aligns with the user's previous purchases of moisturizing products for their baby. It is also fragrance-free, which may be appealing to users who prefer unscented products. \\
    Therefore, based on the user's previous purchases and the product features, I would recommend either the "Bioderma Atoderm Cream for Very Dry or Sensitive Skin" or the "Mustela Stelatopia Cleansing Cream, Baby Body Wash for Extremely Dry to Eczema-Prone Skin, Fragrance-Free, 6.76 Fl. Oz." as the next item for the user to consider purchasing. \\

\bottomrule
\end{tabularx}%
}
\label{tab:reasoning}%
\end{table*}%

\begin{table*}[t]
\centering
\caption{Example of generated retrieval queries and corresponding retrieval results from web search APIs.}
\scalebox{0.9}{
\begin{tabularx}{\textwidth}{X}
    \toprule
    \textbf{Target Item Title:} \\

    \midrule

    Bioderma Atoderm \highlighta{Cream} for Very \highlightb{Dry} or \highlightc{Sensitive Skin} \\

    \midrule
    
    \textbf{Retrieval Query (Our \ours{}):} \\

    \midrule
    
    recommendation product nourishing \highlighta{cream} designed \highlightb{dry} soothing gentle fragrance-free formula \\
    
    \midrule
    
    \textbf{Retrieval Results (Web search API):} \\

    \midrule
    1. Title: ``The Rich Moisturiser Fragrance Free - Skin Rocks''. Content: ``Intensely nourishing, fragrance free rich moisturiser for skin that needs extra care. Addresses \highlightb{dry}ness at its core, leaving skin soft,''. \\
    2. Title: ``Min<u>00e9ral 89 100H Rich \highlighta{Cream} | Vichy Laboratoires''. Content: ``Min<u>00e9ral 89 100H Moisture Boosting Rich \highlighta{Cream} Fragrance- Free provides up to 100 hours of lasting hydration for healthy-looking skin.''. \\
    3. Title: ``The Rich \highlighta{Cream} | Augustinus Bader US''. Content: ``Richly nourishing and loaded with essential fatty acids, including linoleic acid. Effectively reduces moisture loss to soothe, strengthen and enhance skin's''. \\
    4. Title: ``Our Calming \highlighta{Cream} - HealthyBaby''. Content: ``This calming \highlighta{cream} is for \highlightb{dry}, \highlightc{sensitive}, eczema-prone \highlightc{skin}, made with colloidal oatmeal, shea butter, aloe, coconut oil, pre/probiotics, and is fragrance-free.''. \\
    5. Title: ``Revolution Skincare Nourish Boost Rich Nourishing \highlighta{Cream}, Vegan ...''. Content: ``RICH NOURISHING FORMULA: Deep moisturizing \highlighta{cream} designed to provide intense hydration and nourishment to the skin <u>00b7 VEGAN FRIENDLY: Formulated with plant-based''. \\
    6. Title: ``Alcohol and Fragrance-free Moisturizers for \highlightc{Sensitive Skin}''. Content: ``Top picks of Alcohol and Fragrance-free Moisturizers for \highlightc{Sensitive Skin} <u>00b7 COSRX Ultimate Nourishing Rice Overnight Mask <u>00b7 Cerave Moisturizing \highlighta{Cream} <u>00b7 Etude House''. \\
    7. Title: ``8 Best Fragrance-Free Body Lotions for Softer, Less Irritated Skin''. Content: ``Best Overall: La Roche-Posay Lipikar AP+ Triple Repair Moisturizing \highlighta{Cream} <u>00b7 Lipikar AP+ Triple Repair Moisturizing \highlighta{Cream} <u>00b7 Caption Options.''. \\
    8. Title: ``FAVORITE FRAGRANCE FREE PRODUCTS. (\highlightc{SENSITIVE SKIN})''. Content: ``AVEENO Skin Relief Body Wash - <u>00b7 FIRST AID BEAUTY KP BUMP ERASER BODY SCRUB - <u>00b7 NECESSAIRE THE BODY SERUM - <u>00b7 NECESSAIRE THE BODY LOTION - <u>00b7 EUCERIN''. \\
    9. Title: ``Calm + Restore<u>00ae Redness Relief \highlighta{Cream} for \highlightc{Sensitive Skin} | Aveeno<u>00ae''. Content: ``This redness relief face moisturizer formulated with Vitamin B5  ceramide features a calming feverfew  nourishing oat formula to calm \highlightb{dry} skin.''. \\
    10. Title: ``(2 pack) Baby Jergens Soothing \highlightc{Sensitive Skin} \highlighta{Cream}, Newborn ...''. Content: ``This baby \highlighta{cream} for \highlightb{dry} skin softens and soothes your baby's \highlightb{dry} skin, leaving no sticky feel, for 24 hours of hydration. Pediatrician and dermatologist tested,'' \\
\bottomrule
\end{tabularx}%
}
\label{tab:retrieval}%
\end{table*}%

\subsubsection{Hyper-parameter Analysis}
In this section, we delve into the impact of three main hyper-parameters of our proposed \ours{}, as shown in Fig.~\ref{fig:ablation_hyperparameter}, including LLM backbones, top-N retrieval, and top-K recommendations.
Our results reveal that the retrieval backbone exerts a dominant influence on RAG system performance compared to the generation backbone, with Mistral-Instruct achieving substantial gains over LLaMA-Chat up to $6.4\%$ and $5.7\%$ in HR and NDCG metrics, respectively, mainly attributing to a superior instruction-following capacity for recommendation tasks.
These findings demonstrate the critical role of RAG in facilitating the generation performance of LLMs.
Meanwhile, upgrading the generation backbone to LLaMA-2 consistently enhances results over LLaMA, which is consistent with the performance evolution of LLM families.

As for top-N retrieval in Fig.~\ref{fig:topn}, the results demonstrate a clear performance advantage of our \ours{} framework over the vanilla RAG baseline, where website content is naively concatenated as plain text, in top-N website retrieval.
In particular, the performance divergence widens notably at higher N values.
These findings underscore that naively integrating web content, without effective purification or structuring mechanisms, severely constrains recommendation quality.
Notably, our method consistently yields performance improvements as N increases, which demonstrates superior capability in distinguishing relevant information from noisy web data for recommendations.
As illustrated in Fig.~\ref{fig:topk}, we further validate the pattern of performance with respect to top-K recommendations.
In particular, \ours{} demonstrates substantial performance gains over Vanilla RAG, in which such advantages persist at larger K values.
These results validate the effectiveness of our proposed \ours{} for facilitating LLM-based recommendations with web-based RAG, while maintaining scalable ranking quality in recommendation scenarios.

\subsection{Example of Retrieval Results}
In this part, we present an example of the generated retrieval query based on our proposed \ours{}, along with the retrieval results from web search APIs, as shown in Table~\ref{tab:reasoning} and Table~\ref{tab:retrieval}, respectively.
For reference, we take advantage of the reasoning capability of LLMs to interpret recommendation tasks (e.g., \textit{``The recommendation is''}) into specific and informative queries, such as the detailed description of user preferences, that cater to web retrieval.
Specifically, we apply LLMs to generate token sequences, and then sample critical tokens as high-quality retrieval queries based on carefully designed scoring of the LLM information needs in recommendation tasks.
In other words, the retrieval queries are sampled from LLM-generated reasoning on user preferences, which are used to retrieve online websites via API calls.
Notably, we highlight the common keywords between the target item title and the retrieval results from websites, demonstrating the effectiveness of \ours{} to retrieve relevant information in recommendation tasks.
It can be observed that the retrieved web content is highly related to the target item, which contributes to the accurate recommendation of the target item over other candidate items.

%% file: 05conclusion.tex
\section{CONCLUSION}
In this paper, we introduce \ours{}, a novel web-based RAG to facilitate LLM-based recommendations by harnessing up-to-date and high-quality information retrieved from online websites.
Specifically, we tackle the unique challenges due to the significant knowledge gap between web content and recommendation tasks, where noisy web data hardly contributes to the recommendation capability of LLMs.
To bridge this gap, we first take advantage of the reasoning capability of LLMs to interpret recommendation tasks into semantic information on user preferences that cater to web retrieval, sampling LLM-generated tokens as high-quality retrieval queries.
Subsequently, given noisy web-augmented information, where relevant pieces of evidence are scattered far apart, we introduce an attention-guided RAG approach instead of existing matching-based methods.
Drawing inspiration from the retrieval capability of the attention mechanism in LLMs, we design a novel \adapter{} that captures learnable correlations via message passing, enhancing LLM attentions between distant tokens as one-hop connectivity.
Therefore, the long-distance dependencies learned by \adapter{}, which serves as an additional attention head in LLM layers, contribute to capturing relevant information over noisy web content to facilitate recommendation performance. 
Extensive experiments on different real-world recommendation datasets are conducted to demonstrate the effectiveness of our proposed methods under diverse sources of web information.